\documentclass[]{IEEEtran}
\usepackage{mathrsfs}
\usepackage{amsfonts}
\usepackage{graphicx,cite,epsfig,amssymb,amsmath}
\usepackage{color,xcolor}
\definecolor{red}{rgb}{1.00, 0.00, 0.00}
\usepackage{pifont}
\usepackage{stmaryrd}
\usepackage{setspace}
\usepackage{subfigure}
\usepackage{cite}
\usepackage{array}
\usepackage{float}
\usepackage{verbatim}
\usepackage{multirow}
\usepackage[linesnumbered, ruled]{algorithm2e}
\usepackage{epstopdf}
\usepackage{mathtools}
\usepackage{diagbox}
\usepackage{cases}
\usepackage[colorlinks, linkcolor = black, anchorcolor = black, citecolor = black]{hyperref}
\providecommand{\algorithmname}{Algorithm}
\floatname{algorithm}{\protect\algorithmname}
\newcommand{\bm}[1]{\mbox{\boldmath{$#1$}}}
\newtheorem{thm}{Theorem}
\newtheorem{rem}{Remark}
\newtheorem{lem}{Lemma}

\newtheorem{prop}{Proposition}
\usepackage{arydshln}
\allowdisplaybreaks[4]
\newcommand{\revision}[1]{\textcolor{black}{#1}}
\newcommand{\revisionv}[1]{\textcolor{black}{#1}}
\begin{document}

\title{Integrated Sensing, Computation, and Communication: System Framework and Performance Optimization}
\author{\IEEEauthorblockN{Yinghui He, \IEEEmembership{Graduate Student Member,~IEEE},  Guanding Yu, \IEEEmembership{Senior Member,~IEEE}, \\ Yunlong Cai, \IEEEmembership{Senior Member,~IEEE}, and Haiyan Luo}\\	

%\thanks{Manuscript received September 21, 2022; revised May 14, 2023; accepted May 30, 2023. The work of Y. Cai was supported in part by the National Natural Science Foundation of China under Grants 61971376, U22A2004, and 61831004.  (\emph{Corresponding author: G. Yu.})}
\thanks{Y. He, G. Yu, and Y. Cai are with the College of Information Science and Electronic Engineering, Zhejiang University, Hangzhou 310027, China, and also with Zhejiang Provincial Key Laboratory of Information Processing, Communication and Networking (IPCAN), Hangzhou 310027, China (e-mail: 2014hyh@zju.edu.cn;  yuguanding@zju.edu.cn; ylcai@zju.edu.cn).

H. Luo is with Lenovo Research, Shanghai 201210, China (e-mail: luohy7@lenovo.com). }}

%\markboth{IEEE TRANSACTIONS ON WIRELESS COMMUNICATIONS, Vol. XX, No. X, Month 2023}{}

\maketitle
%\vspace{-7ex}
\begin{abstract}
\revision{Integrated sensing, computation, and communication (ISCC) has been recently considered as a promising technique for beyond 5G systems. In ISCC systems, the competition for communication and computation resources between sensing tasks for ambient intelligence and computation tasks from mobile devices becomes an increasingly challenging issue. To address it, we first propose an efficient sensing framework with a novel action detection module. In this module, a threshold is used for detecting whether the sensing target is static and thus the overhead can be reduced. Subsequently, we mathematically analyze the sensing performance of the proposed framework and theoretically prove its effectiveness with the help of the sampling theorem. Based on sensing performance models, we formulate a sensing performance maximization problem while guaranteeing the quality-of-service (QoS) requirements of tasks. To solve it, we propose an optimal resource allocation strategy, in which the minimum resource is allocated to computation tasks, and the rest is devoted to the sensing task. Besides, a threshold selection policy is derived and the results further demonstrate the necessity of the proposed sensing framework. Finally, a real-world test of action recognition tasks based on USRP B210 is conducted to verify the sensing performance analysis. Extensive experiments demonstrate the performance improvement of our proposal by comparing it with some benchmark schemes.}
%Integrated sensing, computation, and communication (ISCC) has been recently considered as a promising technique for beyond 5G systems. In ISCC systems, the competition for communication and computation resources between sensing tasks for ambient intelligence and computation tasks from mobile devices becomes an increasingly challenging issue. To address it, we first propose an efficient sensing framework with a novel action detection module. It can reduce the overhead of computation resource by detecting whether the sensing target is static. Subsequently, we analyze the sensing performance of the proposed framework and theoretically prove its effectiveness with the help of the sampling theorem. Then, we formulate a sensing accuracy maximization problem while guaranteeing the quality-of-service (QoS) requirements of tasks. To solve it, we propose an optimal resource allocation strategy, in which the minimal resource is allocated to computation tasks, and the rest is devoted to sensing tasks. Besides, a threshold selection policy is derived. Compared with the conventional schemes, the results further demonstrate the necessity of the proposed sensing framework. Finally, a real-world test of action recognition tasks based on USRP B210 is conducted to verify the sensing performance analysis, and extensive experiments demonstrate the performance improvement of our proposal by comparing it with some benchmark schemes. 
\end{abstract}

\section{Introduction}
The past decades have witnessed  innovations in network architecture and wireless technology for better communication services, such as high data rate, massive device access, and low latency. On the other hand, with artificial intelligence (AI) showing its power in almost every field, the intelligent revolution has finally come to wireless networks \cite{ML_WC}. Implementing AI in wireless networks can provide diversified intelligent services in cellular networks \cite{intelligent_edge}. One representation among them is to achieve ambient intelligence and smart environments \cite{Ambient} based on a large amount of real-time sensing data. Compared with vision-based sensing, radio-frequency (RF)-based sensing is a more appealing approach for data collection since it can work in non-line-of-sight  scenarios, as well as it is light-needless \cite{wifi_csi_survey} and privacy-preserving \cite{privacy}.  Three key aspects are necessary for enabling RF-based sensing in cellular networks, i.e., spectrum resource, hardware, and computation resource. The former two are used for transmitting and receiving sensing signals, and the last one is used for running sensing algorithms. 

Regarding the first and second aspects, directly allocating extra spectrum and equipping individual hardware would incur high overhead. To address this issue, integrated sensing and communication (ISAC) has been recently proposed since sensing and communication systems have a similar hardware architecture and can work on the same spectrum \cite{ISAC_mag}. \revision{ISAC is expected to improve the spectral and energy efficiencies with a low hardware cost by jointly designing the sensing and communication functions \cite{ISAC_survey}.} To provide computation resource, a novel architecture, referred to mobile edge computing (MEC), has been recently recognized as a core technology by deploying  edge servers with sufficient computation capacity at cellular base stations (BSs) \cite{MEC_survey}. By offloading the computation task to nearby edge servers,  end-to-end latency can be significantly reduced as compared to cloud computing \cite{MEC_survey2}. \revision{Motivated by this,  integrated sensing, computation, and communication (ISCC) has been recently proposed by combining ISAC with MEC to achieve the goal of supporting ambient intelligence and smart environments \cite{SCC_feel,SCC_if}. }

\subsection{Related Work}

\revisionv{ISCC is one of the key technologies for the next generation wireless networks and it will play a vital role in a wide range of application scenarios, such as smart home scenarios \cite{SCC_feel}, internet of vehicle scenarios \cite{iov1,iov2,iov3}, and so on.} ISAC and MEC are fundamental technologies for ISCC. The studies of ISAC have two stages, i.e., spectrum sharing and hardware sharing. For spectrum sharing, radar devices and communication devices share the same spectrum, and the main goal is to improve spectrum efficiency by jointly designing beamforming matrices \cite{ISAC_FS1,ISAC_FS2}. For instance, a joint design of transmit covariance matrix and radar sampling scheme has been proposed in \cite{ISAC_FS1} to reduce the effective interference power at the radar receiver. Moreover, since the sensing and communication systems have a similar hardware design, hardware sharing becomes possible to reduce the overhead. One way to realize hardware sharing is to enable the sensing function within a communication system \cite{isac_h1,isac_h2,isac_h3}. For example, the work in \cite{isac_h1} utilized a commercial WiFi card with a modified driver to realize the localization function. \revision{In addition, the other method is to design a dual-functional waveform that further exploits the potential of ISAC \cite{isac_wave1,isac_wave2,isac_wave3,isac_wave4}. For instance, the authors of \cite{isac_wave1} proposed an optimal waveform toward a trade-off between radar and communication performance. }

However, the above works only focused on the processes of transmitting and receiving sensing signals and have not thoroughly studied the computation process of sensing data. Note that for delay-sensitive sensing tasks, the computational delay is non-negligible. Meanwhile, offloading sensing data to cloud servers certainly incurs high communication latency. MEC can be adopted to address this issue since it provides computation ability to the edge network by equipping BSs with edge servers. The studies of MEC mainly concentrate on joint communication and computation resource allocation for minimizing the total energy consumption  \cite{mec1,mec2} or end-to-end delay \cite{mec3,mec4}. 

Nevertheless, the design of resource allocation in MEC has not considered sensing tasks and their features. As a result, a major issue for ISCC is how to allocate the limited available resources, taking into account both communication and computation. \revision{Existing works \cite{SCC_irs,SCC_feel,SCC_if,SCC_mimo} have proposed their solutions for different scenarios to address this problem.  The authors in \cite{SCC_mimo} considered a general sensing task and jointly optimized computation resource allocation and MIMO precoding for radar waveforms and communication symbols with the delay constraint and resource limitation in ISCC systems. To reduce the interference between sensing and communication, intelligent reflecting surface (IRS) is adopted in \cite{SCC_irs} to suppress the interference between sensing and communication. A joint resource allocation and beamforming algorithm is proposed to minimize energy consumption. Differently, literatures \cite{SCC_feel,SCC_if} investigated AI-based sensing tasks.} Specifically, the training process of human motion recognition tasks in a federated edge learning (FEEL) system has been investigated in \cite{SCC_feel}, and the resource allocation problem of ISCC was addressed to accelerate the convergence of FEEL. On the other hand, the authors of \cite{SCC_if} studied the inference process of AI-based sensing tasks where the sensing accuracy is measured by an approximate but tractable metric, i.e., discriminant gain. An optimal joint transmit power and communication resource allocation strategy has been proposed for maximizing the discriminant gain. However, directly implementing AI-based sensing algorithms will bring a high computational expense in ISCC systems since they are computation-intensive. Meanwhile, in most scenarios, such as smart home scenarios, the sensing target might be static for an extended period\footnote{\revision{The static state means that the sensing scenario remains unchanged. For example, a human (i.e., the sensing target) keeps the same posture, which would not cause the variations of sensing signals over time.}}, or there may even be no sensing target. \revision{Besides, existing works about RF-based sensing mainly focus on generalization issue \cite{generalization1,generalization2,generalization3} and privacy issue \cite{privacy,privacy1,privacy2}. The former tries to improve the generalization ability of the sensing algorithm for automatically adapting to new and previously unseen sensing targets and environments, since RF signals are sensitive to the environments and sensing targets. The latter one aims to avoid the information leakage, since RF signals are non-intrusive and can be transmitted through the wall. Nevertheless, the computation overhead issue of sensing tasks has not been studied before.} Inspired by this, we aim to propose a novel sensing framework to further improve computational and communication efficiencies while guaranteeing sensing performance.

\subsection{Main Contributions}
In this paper, we consider an ISCC system where a triple-functional BS equipped with an edge server needs to perform a sensing task, i.e., action recognition, for the ambient intelligence and also provide communication and computing services in the cellular network. \revision{Specifically, the sensing task serves the smart home scenarios where variations of wireless signals could be employed to detect or recognize human daily activities and thereby the BS can support diversified intelligent services, such as fall detection and intelligent home control \cite{ISAC_mag}. Since the sensing and computation tasks compete for the limited computation and communication resources, it is crucial to design an effective framework and resource allocation strategy.} This work is mostly related to the prominent ISAC work \cite{ISAC_learning}. It only focuses on the tradeoff between recognition accuracy and communication rate. Moreover, a convolutional neural network (CNN)-based sensing algorithm was directly used in \cite{ISAC_learning} for sensing tasks without considering the computation resource waste for the static state. Motivated by this, we aim to design an effective sensing framework and the corresponding resource allocation strategy to further improve the sensing, computation, and communication performance simultaneously. In particular, we address three main challenges: 1) how to design this sensing framework; 2) how to analyze the sensing performance based on the proposed sensing framework; 3) how to jointly allocate resources for maximizing the sensing performance under the quality-of-service (QoS) requirements of devices. The main contributions of this work are summarized as follows.
\begin{itemize}
\item We propose a novel and effective sensing framework where a key action detection module is placed before the CNN module. In the action detection module, the power of high-frequency components of sensing signals is compared with a threshold for determining whether the sensing target is static due to the fact that the action of the target will cause the variations of sensing signals.
\item We quantitatively analyze the overall sensing performance of the proposed sensing framework, i.e., the sensing accuracy and delay. Specifically, the sampling theorem is applied to calculate the miss rate and false positive rate for the action detection module. The accuracy is modeled as a monotonically increasing function, and the delay of the CNN module is proportional to the sampling rate. Furthermore, we prove that our proposed sensing framework can save unnecessary computation resource, and the gain is positively related to the probability of the occurrence of the static state.
\item A sensing accuracy maximization problem is formulated with the delay requirement of computation tasks. By addressing this problem, we propose an optimal resource allocation strategy in which the resource allocated to computation tasks is minimized and the remaining resource is allocated to the sensing task to maximize the accuracy. Besides, an optimal threshold selection policy is also derived. 
\item We conduct a real-world test of the action recognition task using the USRP B210. The proposed mathematical models are validated with the collected dataset. Furthermore, numerical results demonstrate that the proposed sensing framework and algorithm outperform several other benchmark schemes.
\end{itemize}

The rest of the paper is organized as follows. Section II introduces the ISCC system model, the computation task model, and the sensing framework. Section III analyzes the sensing performance of the proposed framework and proves its superiority. Section IV formulates a sensing performance maximization problem and decomposes it into two subproblems for developing an overall algorithm. Section V presents the test results, and the whole paper is concluded in Section VI.

\section{System Model and Sensing Framework}

In this section, we introduce the ISCC system, the computation task model, and the proposed effective sensing framework.
\subsection{ISCC System}

\begin{figure}[t]
	\vspace{-3ex}
	\centering
		%\vspace{-2ex}
		\includegraphics[width=0.9\linewidth]{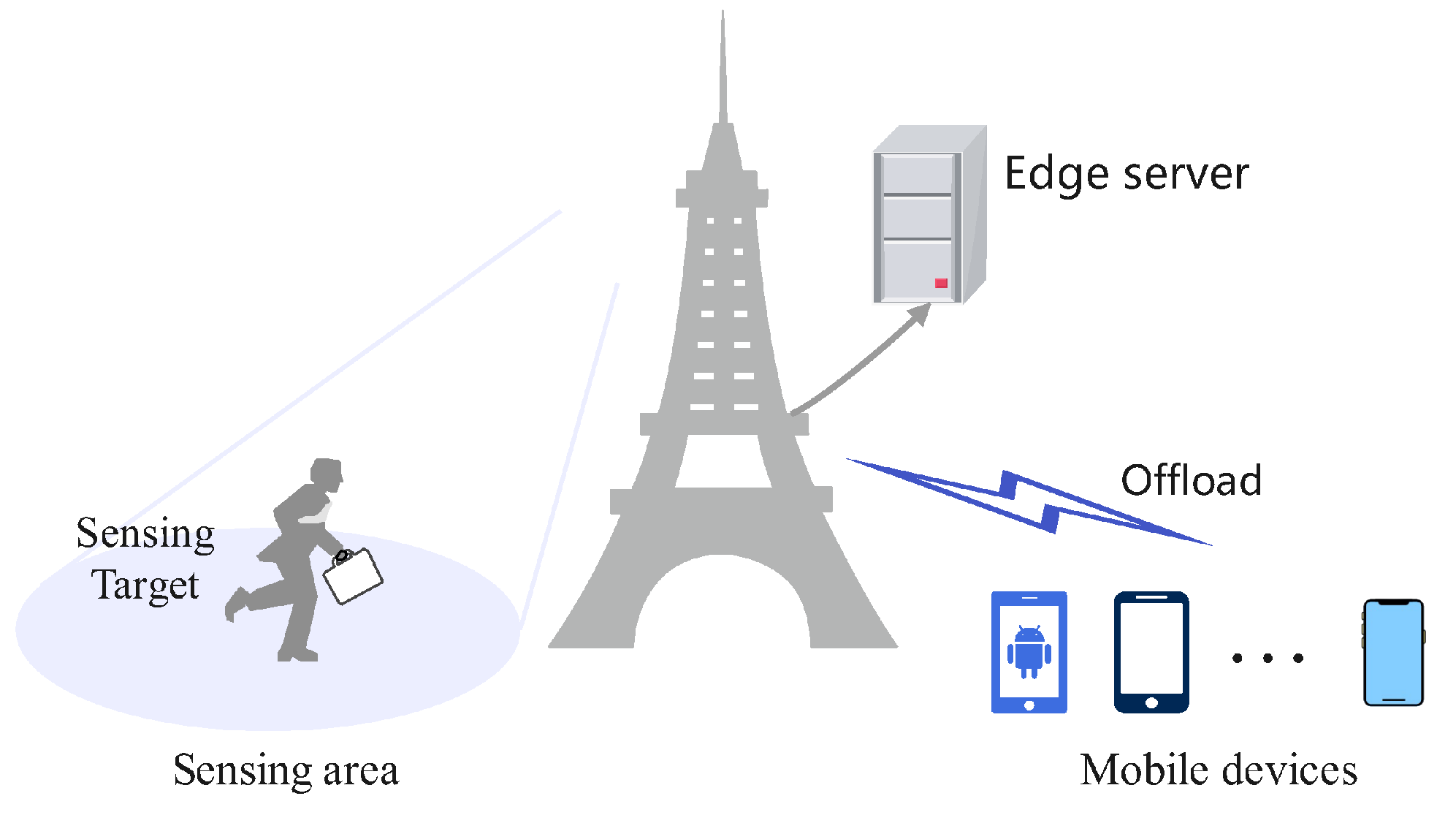}
		\vspace{-2ex}
		\caption{An ISCC system.}
		\vspace{-2ex}
		\label{fig:sys}
\end{figure}

As depicted in Fig. \ref{fig:sys}, we consider an ISCC system containing one triple-functional BS equipped with an edge server and $N$ mobile devices. The BS aims to conduct a sensing task and provide communication and computing services. Each device has its own computation task. Due to the limited computation resource at the mobile device, all computation tasks need to be offloaded to the edge server located at the BS via the wireless links.  Meanwhile, the BS needs to perform a sensing task, i.e., an action recognition task.  For this purpose,  the wireless sensing signal is transmitted towards the sensing area at the sampling rate $F^{\text{s}}$. The echo is collected at the triple-functional BS and then fed back to the edge server for action recognition. \revision{To sum up, the BS has three functions: 1) communication function: mobile devices can offload their tasks to the BS via wireless links; 2) sensing function: the BS can transmit sensing signals; 3) computation function: the computation and sensing tasks can be completed with the help of the edge server.}

\begin{figure}[htbp]
	\vspace{-3ex}
	\centering
	%\vspace{-2ex}
	\includegraphics[width=0.83\linewidth]{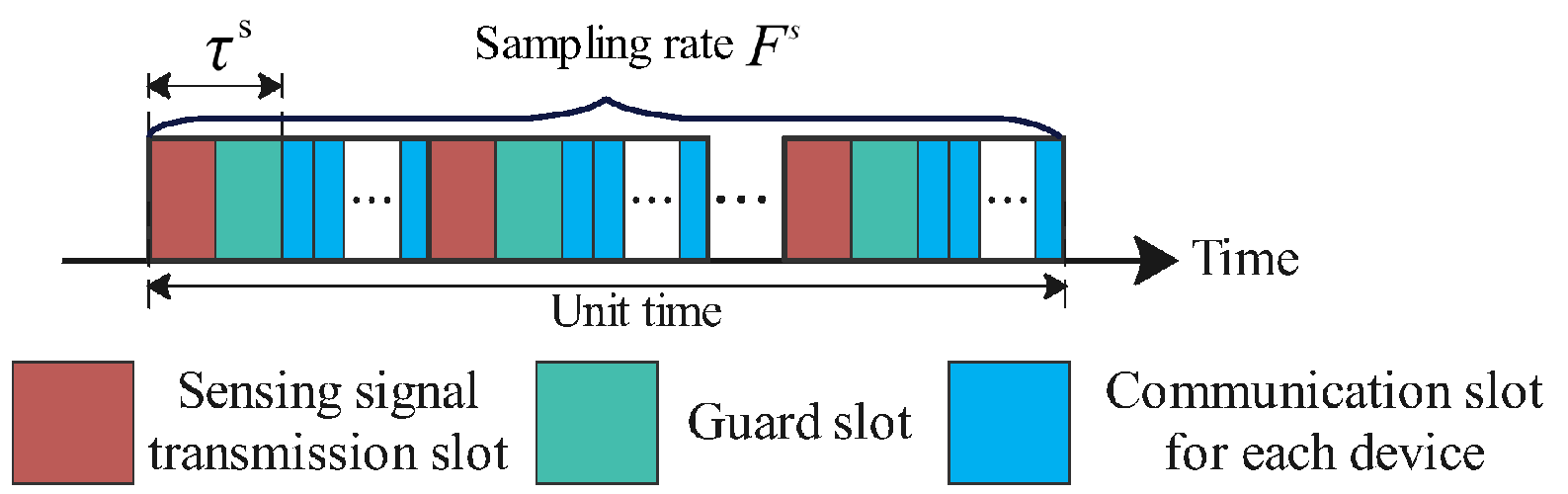}
	\vspace{-1ex}
	\caption{\revision{Illustration of TDMA method at the triple-functional BS.}}
	\label{fig:time}
	\vspace{-2ex}
\end{figure}

\revision{In this work, since the interference between sensing and communication would seriously affect the sensing performance, the BS adopts the time division multiple access (TDMA) method to perform sensing and communication functions, as shown in Fig. \ref{fig:time}.\footnote{\revision{We should note that the whole system is under the control of the BS and each device only can transmit data within the allocated time slots.}}} Specifically, there are three types of time slots: 1) sensing signal transmission slot for transmitting the sensing signal; 2) guard slot for collecting sensing echoes; 3) communication slot for data offloading. \revision{Let $\tau^{\text{s}}$ denote the total proportion of one sensing signal transmission slot and one guard slot, i.e., total duration per unit time. Moreover, the BS transmits sensing signals at a sampling rate of $F^{\text{s}}$, which means there are $F^{\text{s}}$ sensing signal transmission slots and  $F^{\text{s}}$ guard slots per unit time. Thus, the total proportion of communication resource occupied  by the sensing task is $\tau^{\text{s}}F^{\text{s}}$.  The proportion of communication resource allocated to device $n$ for offloading per unit time is denoted by  $\tau_n^{\text{c}}$.\footnote{\revision{The delay of computation tasks is not related to the sampling rate, and thus we directly describe the proportion of communication resource allocated to device $n$ as $\tau_n^{\text{c}}$.}} Therefore, we have $\tau^{\text{s}}F^{\text{s}} + \sum_{n=1}^N \tau^{\text{c}}_n \le 1$.}\footnote{\revisionv{Compared with the duration of the time slot for sensing and communication tasks, the duration of the controlling signal is much shorter and thus can be ignored. This assumption has been widely adopted in similar works \cite{SCC_feel, SCC_if}.}}

%$\tau^{\text{s}}$ represents the total proportion of one sensing signal transmission slot and one guard slot, i.e., total duration per unit time. Moreover, the BS transmits sensing signal with  the sampling rate being $F^{\text{s}}$, which means there are $F^{\text{s}}$ sensing signal transmission slots and  $F^{\text{s}}$ guard slots per unit time. Thus, the total proportion of communication resource occupied  by the sensing task is $\tau^{\text{s}}F^{\text{s}}$. 

\subsection{Computation Task Model}
The computation task of device $n$ can be represented by a tuple $(V_n, C_n, T_n^{{\max}})$, where $V_n$ (in bit) denotes the data size of the task, $C_n$ (in CPU cycle/bit) denotes the computation intensity, i.e., the required number of CPU cycles for computing one bit, and $T_n^{{\max}}$ (in s)  denotes the delay limitation. The computation task offloading of each device can be divided into two main phases: 1) \emph{transmission phase}, where each device transmits its task to the BS with the allocated time slots; 2) \emph{computation phase}, where the edge server computes the task. Note that the phase of returning the computation result can be neglected because of the small data size \cite{mec1}.  

In the transmission phase, let $B$ denote the system bandwidth and let $\sigma^2_\text{c}$ denote the channel noise power. Then, the instantaneous data rate of device $n$ is given by 
\begin{equation}
	R_n = B \log_2\left(1+ \frac{|h_n|^2 p_n}{\sigma^2_\text{c}}\right),
\end{equation}
where $h_n$ is the channel gain between device $n$ and the BS, and $p_n$ is the transmit power of device $n$. \revision{Meanwhile,  the proportion of communication resource occupied by device $n$ is  $\tau^\text{c}_n$, that is, the time slot available for transmission of device $n$ is $\tau_n$ per unit time. Thus, for device $n$, the transmitted data size per unit time is $\tau_n^c R_n$, which means that the average data rate of device $n$ is ${\tau}_n^cR_n$.} Thus, for device $n$, the delay of  transmission phase can be expressed as
\begin{equation}
	T^{\text{tran}}_n = \frac{V_n}{\tau_n^\text{c} R_n}.
\end{equation}

In the computation phase, let $f_n$ (in CPU cycle/s, i.e., Hz) denote the computation resource of the edge server allocated to device $n$. Then, according to \cite{MEC_model}, the delay of  computation phase for device $n$ can be expressed as
\begin{equation}
	T^{\text{comp}}_n = \frac{V_nC_n}{f_n}.
\end{equation}
Based on the above, the overall delay for device $n$ to complete its task is given by 
\begin{equation}
	T_n = T^{\text{tran}}_n + T^{\text{comp}}_n = \frac{V_n}{\tau_n^\text{c} R_n} + \frac{V_nC_n}{f_n}. \label{eq:T_device}
\end{equation}

\subsection{Sensing Framework} \label{subsec:sensing_frame}
\begin{figure}[h]
	\vspace{-2ex}
	\centering
	\includegraphics[width=1\linewidth]{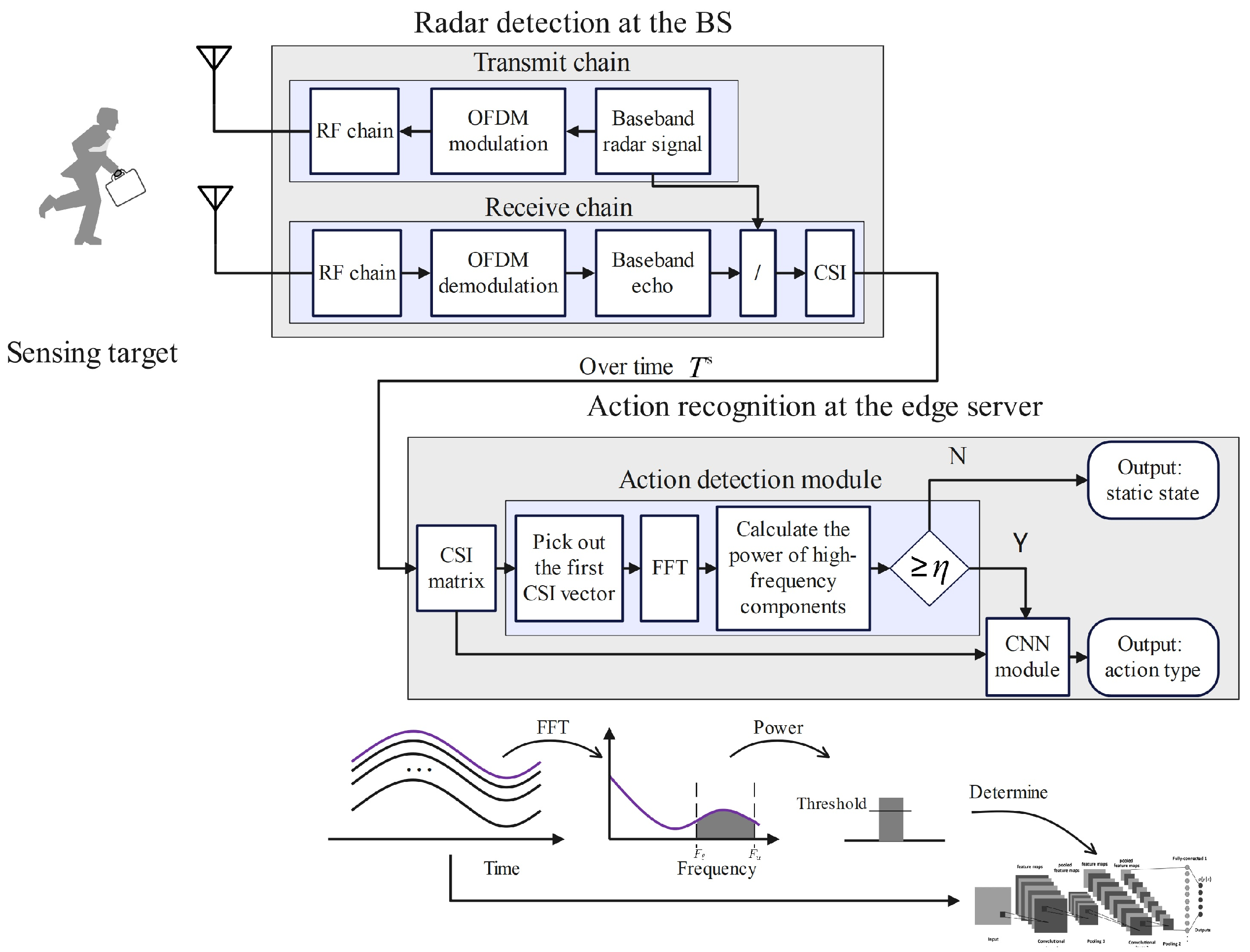}
	\vspace{-3ex}
	\caption{\revision{The proposed sensing framework.}}
	\label{fig:flow}
\end{figure}

For the sensing task of action recognition, the overall sensing framework contains two phases, i.e., \emph{signal collection phase} at the BS and \emph{action recognition phase} at the edge server, as shown in Fig. \ref{fig:flow}. The former one is utilized for collecting the radar echo signal, and the latter one is utilized for processing the signal to obtain the action type.\footnote{In this paper, we adopt the CNN-based sensing algorithm since it is one of the most popular solutions for the RF-based action recognition task \cite{CNN_sensing}. Moreover, our proposed method can also be extended to other sensing algorithms.}

For the \emph{signal collection phase} at the BS, we consider an orthogonal frequency division multiplexing (OFDM) radar \cite{ofdm_radar} because it can be easily implemented in existing BSs.\footnote{Our design can be extended to other radar types, such as frequency-modulated continuous wave (FMCW) radar \cite{fmcw}.} With the OFDM radar, the total system bandwidth $B$ is divided into $K$ subcarriers. The baseband radar signal, denoted by $x_k$ for subcarrier $k$, is transmitted via an RF chain after OFDM modulation. After reflection and OFDM demodulation, the baseband echo, denoted by $y_k$ for subcarrier $k$, is obtained. Finally, the processed OFDM radar signal for subcarrier $k$ can be calculated by $g_k = \dfrac{y_k}{x_k}$, which can be regarded as the channel state information (CSI) between the transmitting and receiving antennas of the BS. \revision{To detect the action type of sensing target, we need to collect the CSI samples over a period of time, whose length is denoted by $T^{\text{s}}$.} With the sampling rate being $F^{\text{s}}$, the total collected CSI samples can construct a CSI matrix, denoted by $\bm{G} =[\bm{g}_1^H~\bm{g}_2 ^H \cdots \bm{g}_K ^H]^H \in \mathbb{C}^{K\times( T^{\text{s}}F^{\text{s}})}$, where $\bm{g}_k =[g_{k}[1]	~g_{k}[2]\cdots g_{k}[T^{\text{s}}F^{\text{s}}]]\in \mathbb{C}^{1\times ( T^{\text{s}}F^{\text{s}})}$ is the collected CSI vector for subcarrier $k$ over the period of $T^{\text{s}}$,  $g_{k}[m]$ is the $m$-th CSI sample for subcarrier $k$, and $(\cdot)^H$ denotes the operation of conjugate transpose.

For the \emph{action recognition phase} at the edge server, in the conventional scheme, the collected CSI matrix is directly input into the CNN module \cite{conv1}. However, in some scenarios (especially smart home scenarios),  the target may not exist in the sensing area, or the sensing target may be in a static state. Thus, continuously employing CNN would result in high cost of computation resource. Meanwhile, we should note the fact that an action of a sensing target  causes the CSI to vary over time, which leads to an increase in the power of high-frequency components. Motivated by this, we design an action detection module placed before the CNN module, as shown in Fig. \ref{fig:flow}. It is used for detecting whether the sensing target is in a static state or an action state. To achieve it, we first pick the CSI vector of the first subcarrier among the $K$ subcarriers.\footnote{Here, we take the first subcarrier as an example. In fact, any of the $K$ subcarriers can be used for the action detection module. \revision{Moreover, all $K$ subcarriers can be used in the proposed action detection module with the help of a voting classifier, which deserves future work.}} Then,  to obtain the power of high-frequency components, we need to convert the CSI vector from the time domain to the frequency domain. Thus, we adopt the fast Fourier transform (FFT) to obtain the discrete Fourier transform (DFT) of the CSI vector, as
\begin{equation}
	D[l] = \frac{1}{\sqrt{T^{\text{s}}F^{\text{s}}}}\sum_{m=1}^{T^{\text{s}}F^{\text{s}}}g_{1}[m] \exp\left(\frac{-j 2\pi lm}{T^{\text{s}}F^{\text{s}}}\right).
\end{equation}
Next, we need to calculate the power of high-frequency components.  Let  $[F_{\ell}, F_u]$ denote the range of high frequency and then the corresponding range for DFT $D[l]$ is $\left[\lfloor  F_{\ell}T^{\text{s}} \rfloor,  \lceil F_uT^{\text{s}} \rceil \right]$ according to \cite{signal_system}. The power of  high-frequency components can be calculated as the ratio of the total energy in the range $[F_{\ell}T^{\text{s}}, F_uT^{\text{s}}]$ to the number of samples, as
\begin{equation}
		P = \frac{1}{T^{\text{s}} F^{\text{s}}}  \sum_{l= \lfloor  F_{\ell}T^{\text{s}} \rfloor }^{\lceil F_uT^{\text{s}} \rceil} \left| D[l] \right|^2.
\end{equation}
Finally, by comparing the power $P$ with a pre-set threshold, denote by $\eta$, we can determine whether the sensing target is in the static state or the action state. If it is in the action state, the CSI matrix $\bm{G}$ is further input into the CNN module for action recognition; otherwise, the result of the static state is output. 

The key to the proposed action detection module is the threshold. If the threshold is set too high, it may cause a large number of action state instances to be recognized as the static state, which pulls down the average recognition accuracy. If the threshold is set too low, a number of static state instances may be detected as the action state, increasing both the cost of computation resource and delay. Therefore, it is important to select an appropriate threshold to reduce both computation consumption and delay without  sacrificing accuracy. Moreover, the sampling rate also influences the accuracy. With a higher sampling rate,  it can intuitively be seen that the accuracy becomes higher since more information about the sensing target is collected. In the next section, we will analyze the effect of threshold and sampling rate on the average accuracy and delay of the proposed sensing framework.

\section{Sensing Performance Analysis}
In this section, we first analyze the sensing accuracy and delay of our proposed sensing framework and then prove its effectiveness.
\subsection{Sensing Accuracy} 
In our proposed sensing framework, the sensing accuracy is jointly determined by two modules, i.e., the action detection module and the CNN module. We assume that the action recognition task has $I$ recognition types and their set is denoted by $\{1,2,\cdots,I\}$, where the first type is the static state and $\{2,3,\cdots,I\}$ is the set of action state types.\footnote{\revision{The number of types, i.e., $I$, should be given in the prior design and training processes for AI-based sensing tasks. The value of $I$ is generally determined according to specific scenarios and tasks. In the test part, we consider 8 typical action types.}} For the action detection module, we can describe its performance in terms of miss rate and false positive rate. To be specific, the miss rate for the $i$-th action type is defined as the proportion of the $i$-th action instances that are detected as the static state, i.e., 
\begin{equation}
	p^o_i = \frac{M_i^0}{M_i},~i=2,3,\cdots,I,
\end{equation}
where $M_i$ is the total number of the $i$-th action type instances and $M_i^0$ is the number of the $i$-th action type instances that are recognized as the static state. The false positive rate is defined as the proportion of the static state instances that are recognized as the action state, i.e., 
\begin{equation}
p^l =  \frac{M_0^\text{a}}{M_0},
\end{equation}
where $M_0$ is the total number of static state instances and $M_0^\text{a}$ is the number of static state instances that are recognized as the action state.  For the CNN module, recognition accuracy is utilized to describe the performance, which is defined as the ratio of the correctly recognized instances to the total  number of recognition instances. \revision{Although we have defined performance metrics, they cannot be used for theoretical analysis since they are based on statistical results. Thus, in the following, we will theoretically model them with the help of the sampling theorem.}

\begin{figure}[t]
	\vspace{-1ex}
	\centering
    	\subfigure[The sampling rate is $50$ Hz.]{
        		\centering
        		\includegraphics[width=0.35 \textwidth]{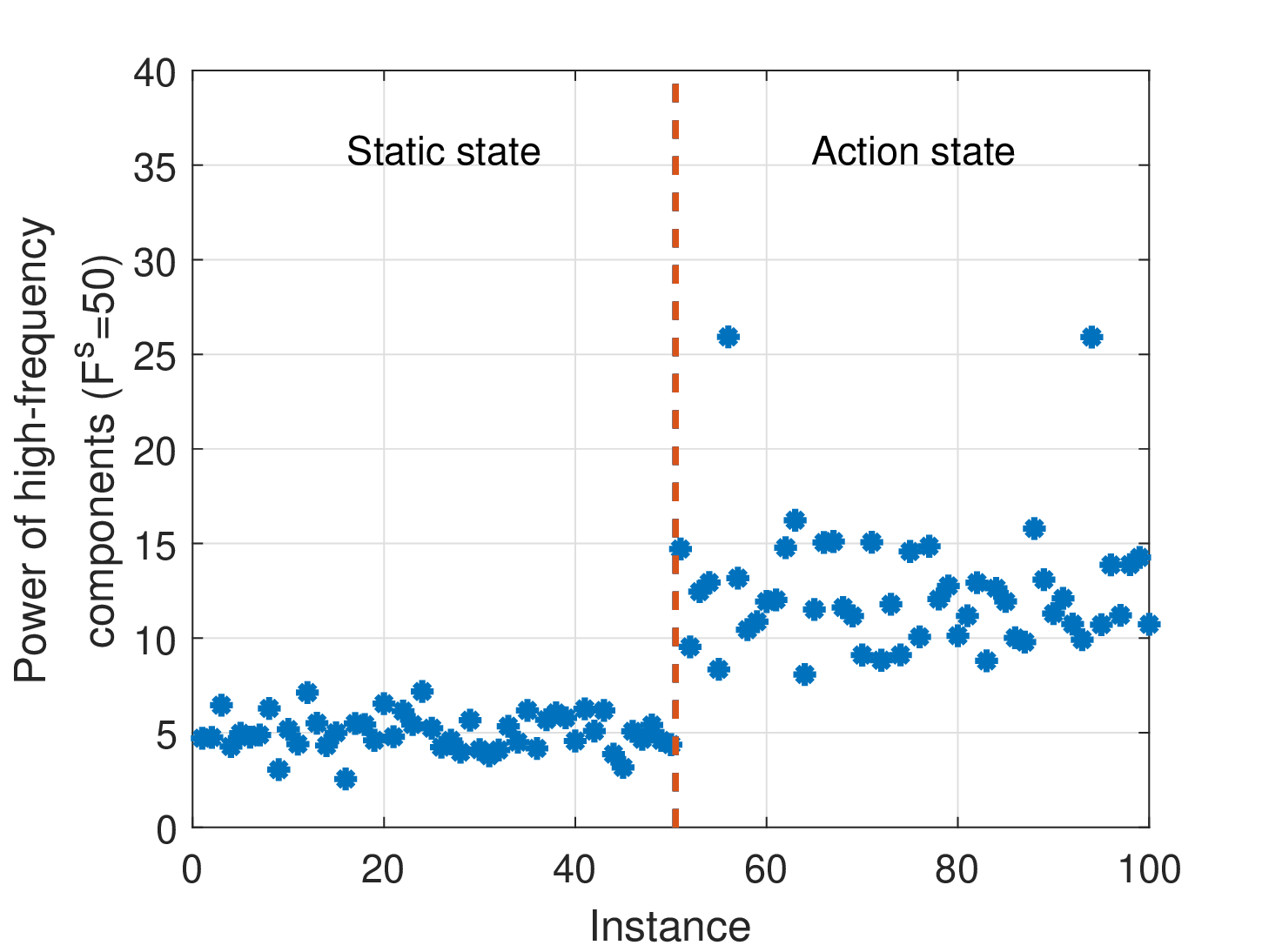}
        	}
    	\subfigure[The sampling rate is $1000$ Hz.]{
        		\centering
        		\includegraphics[width=0.35 \textwidth]{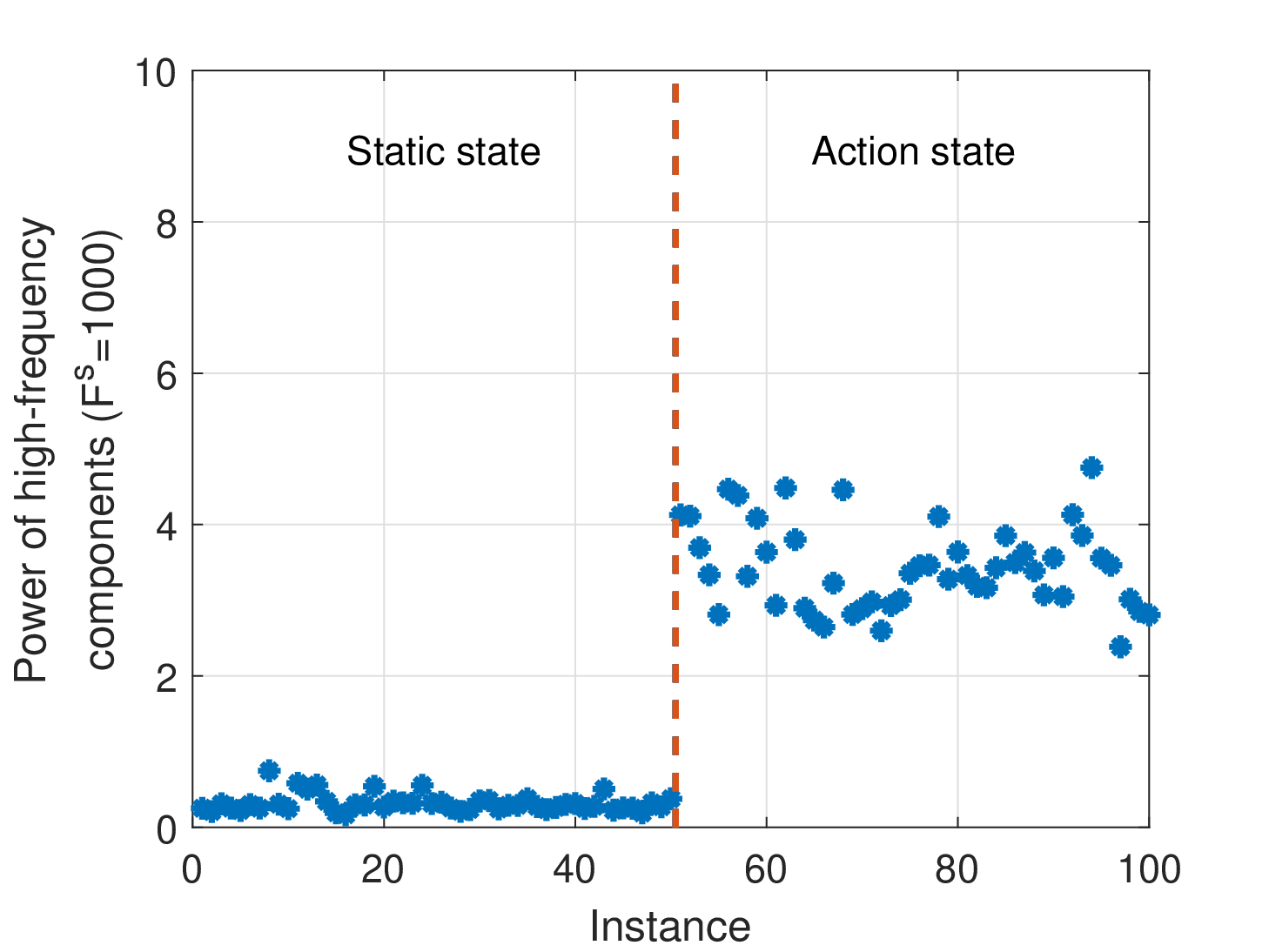}
        	}
	\vspace{-1ex}
	\caption{The power of high-frequency components for 100 instances under different sampling rates. Specifically, for the first 50 instances, the target is in the static state. For the last 50 instances, the target is in the action state. \revision{The test setting and dataset are introduced in Section \ref{sec:sim_para}.}}
	\vspace{-3ex}
	\label{fig:power_example}
\end{figure}

First of all, we focus on the action detection module.  As we mentioned before, the key is the threshold. Moreover, the sampling rate also affects $p^o_i$ and $p^l$. To illustrate the effects of the threshold and sampling rate, we collect some instances and plot the power of high-frequency components in Fig. \ref{fig:power_example}. From the figure, we can observe that the power of the action state is obviously higher than that of the static state, which demonstrates the effectiveness of the proposed action detection module. Besides,  as the sampling rate increases, the power gap between the two states becomes higher, which further reduces both the miss rate and the false positive rate. It is intuitively reasonable since the sensing target becomes more informative, and it is easier to distinguish between the static state and the action state with a higher sampling rate. Recall that we aim to optimize the sensing performance in this paper.  The challenge, however, is that the mathematical relationships among the sampling rate, threshold, miss rate, and false positive rate are unclear. 

To model the effect of sampling rate, we need to consider the continuous-time CSI of the first subcarrier in the time domain, denoted by $g^c_{1}(t)$. Then, the collected CSI can be rewritten as the sample of the continuous-time CSI at sampling rate $F^{\text{s}}$, as
\begin{align}
g_{1}[m] &= g^{\text{c}}_{1}(t)|_{t=m/F^{\text{s}}} + w_0[m] \\
& =g^{\text{c}}_1 \left(\frac{m}{F^{\text{s}}}\right)+ w_0[m], ~m=  1,2,\cdots,T^{\text{s}}F^{\text{s}},
\end{align}
where $w_0[m]$ is the estimation error that can be modeled as the Gaussian noise with mean being zero and variance being $\sigma^2$. Then, according to the sampling theorem \cite{signal_system}, the relationship between the  DFT $D[l]$ of $g_{1}[m]$  and the Fourier transform of $g^{\text{c}}_{1}(t)$, denoted by $D^{\text{c}}(F)$, can be expressed as
\begin{align}
	D[l] &= \frac{1}{\sqrt{T^{\text{s}}F^{\text{s}}}} \left(\sum_{m=0}^{T^{\text{s}}F^{\text{s}}-1}  g^{\text{c}}_{1} \left(\frac{m}{F^{\text{s}}}\right)\exp\left(\frac{-j 2\pi lm}{T^{\text{s}}F^{\text{s}}}\right)\right. \\ \nonumber 
	&\left.~ ~~~~~~~~~~~~~~~+  \sum_{m=0}^{T^{\text{s}}F^{\text{s}}-1}  w_0[m] \exp\left(\frac{-j 2\pi lm}{T^{\text{s}}F^{\text{s}}}\right) \right)  \\
	&= \sqrt{\frac{F^{\text{s}}}{T^{\text{s}}}}  \sum_{m=-\infty}^{+\infty}D^{\text{c}}\left(   \left(\frac{l}{T^{\text{s}}F^{\text{s}}}-m\right)F^{\text{s}}\right) +   W_0[l],\label{eq:DFT}
\end{align}
where $W_0[l]$ is the equivalent Gaussian noise in the frequency domain with zero mean and variance $ \sigma^2$. Supposing $D^{\text{c}}(F)=0$ when  $|F|>{F^{\text{s}}}/{2}$, we can further simplify (\ref{eq:DFT}) as 
\begin{equation}
D[l]=\sqrt{\frac{F^{\text{s}}}{T^{\text{s}}}}  D^{\text{c}}\left(   \frac{  l}{T^{\text{s}}}\right) + W_0[l].
\end{equation}
Then,   the power calculated in the proposed action detection module can be rewritten as 
\begin{align}
	P &=  \frac{1}{T^{\text{s}} F^{\text{s}}}  \sum_{l= \lfloor  F_{\ell}T^{\text{s}} \rfloor }^{\lceil F_uT^{\text{s}} \rceil} \left|  \sqrt{\frac{F^{\text{s}}}{T^{\text{s}}}}  D^{\text{c}}\left(  \frac{  l}{T^{\text{s}}}\right) + W_0[l]\right|^2  \\
	& =  \frac{1}{\left( T^{\text{s}} \right)^2}  \sum_{l= \lfloor  F_{\ell}T^{\text{s}} \rfloor }^{\lceil F_uT^{\text{s}} \rceil} \left|   D^{\text{c}}\left(   \frac{  l}{T^{\text{s}}}\right) +  \sqrt{\frac{T^{\text{s}}}{F^{\text{s}}}}  W_0[l]\right|^2.
\end{align} 
Thus,  $P$ follows the noncentral chi-squared distribution with the degree of freedom (DoF) of $ \lceil F_uT^{\text{s}} \rceil -\lfloor  F_{\ell}T^{\text{s}} \rfloor  +1$. Since the high frequency range is wide enough and $T^{\text{s}}$ is long enough, e.g., $F_{\ell}=10$ Hz and $T^{\text{s}} = 3$ s, $P$ approximately follows the normal distribution according to \cite{chi_square}  and its mean and variance  are given by 
\begin{equation}
	\mu_P =  \lambda + \frac{r}{T^{\text{s}} F^{\text{s}}}, ~\sigma^2_P = 4 \frac{\sigma^2}{T^{\text{s}}F^{\text{s}}}\lambda  + 2 \frac{\sigma^2 }{\left( T^{\text{s}} F^{\text{s}}\right)^2}r, \label{eq:mu}
\end{equation}
respectively, where
\begin{equation}
\begin{gathered}
	\lambda = \frac{1}{\left(T^{\text{s}}\right)^2} \sum_{l= \lfloor  F_{\ell}T^{\text{s}} \rfloor }^{\lceil F_uT^{\text{s}} \rceil} \left|   D^{\text{c}}\left(  \frac{ l}{T^{\text{s}}}\right) \right|^2,\\
	r = \sigma^2 \left( \lceil F_uT^{\text{s}} \rceil -\lfloor  F_{\ell}T^{\text{s}} \rfloor  +1 \right).
 \label{eq:lambda_P}
\end{gathered}
\end{equation}
Moreover, instances of the same action cannot be exactly the same. For example, the target may not stand at the same location when the target performs the same action. Therefore, we add a constant, denoted by $\sigma_{d}^2$, into the variance and it describes the distinction of action instances. Therefore, the revised variance is given by
\begin{equation}
	\sigma^2_P = 4 \frac{\sigma^2}{T^{\text{s}}F^{\text{s}}}\lambda + 2 \frac{\sigma^2 }{\left( T^{\text{s}} F^{\text{s}}\right)^2}r + \sigma_{d}^2. \label{eq:sigma}
\end{equation}
Based on the above analysis, we can mathematically model the relationship among the sampling rate, threshold,  miss rate, and false positive rate, as shown in the following proposition.
\begin{prop}\label{prop:rate}
	The miss rate for the $i$-th action is given by
\begin{equation}
	p^o_i  = Q\left(\dfrac{\mu_{P,i} - \eta}{\sigma_{P,i}}\right), i=2,3,\cdots,I,\label{eq:p_o}
\end{equation}
and the false positive rate for the static state is given by
\begin{equation}
	p^l = Q\left(\frac{\eta - \mu_{P,1}}{\sigma_{P,1}}\right), 
	\label{eq:p_l}
\end{equation}
where $Q(\cdot)$ is the Q-function\footnote{https://cnx.org/contents/hDU5uzaA@2/The-Q-function}, $\mu_{P,i} =  \lambda_{i} + \dfrac{r_{i}}{T^{\text{s}} F^{\text{s}}}$, and  $\sigma_{P,i}^2 =  4 \dfrac{\sigma^2}{T^{\text{s}}F^{\text{s}}}\lambda_{i}  + 2 \dfrac{\sigma^2 }{\left( T^{\text{s}} F^{\text{s}}\right)^2}r_{i}+\sigma_{d,i}^2$.  $\lambda_{i}$, $r_{i}$, and $\sigma_{d,i}^2$ are parameters that can be obtained via fitting on the collected sensing dataset.  
\end{prop}
 
From Proposition \ref{prop:rate},  we can find that $p^o_i$ increases but $p^l$ decreases with the threshold, which confirms the statement we mentioned in Section \ref{subsec:sensing_frame}. Moreover, both the mean and variance decrease with sampling rate, which is consistent with Fig. \ref{fig:power_example}.

\begin{figure}[t]
	\vspace{-4ex}
	\centering
	\includegraphics[width=0.35\textwidth]{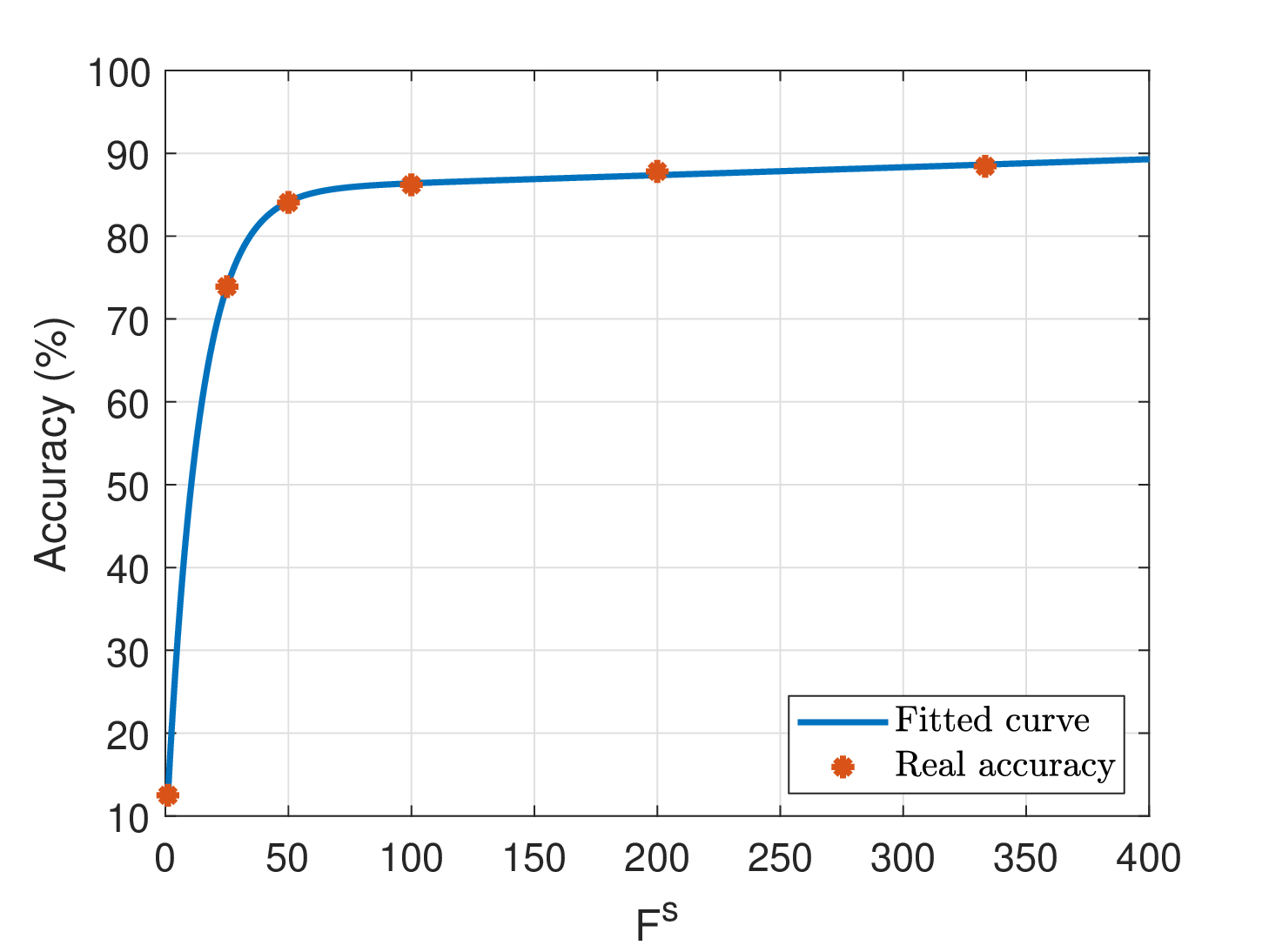}
	\vspace{-3ex}
	\caption{The relationship between the recognition accuracy and sampling rate.}
	\label{fig:acc}
	\vspace{-4ex}
\end{figure}

Next, we need to model the accuracy of the CNN module.  According to \cite{ISAC_learning}, the accuracy is positively correlated with the sampling rate. A higher sampling rate means more information about the sensing target, and it is reasonable that the CNN module can achieve higher accuracy. Thus, it can be represented by a monotonically increasing function, denoted by $\alpha(F^{\text{s}})$.  To verify this, we  collect a sensing dataset and adopt  one of the most well-known CNNs,  ResNet50 \cite{resnet}. After fully training until convergence, the relationship between the accuracy and sampling rate is shown in Fig. \ref{fig:acc}. We can observe that the accuracy is monotonically increasing with $F^{\text{s}}$. 

Based on the above analysis of two modules, we can express the recognition accuracy of each type and show the average recognition accuracy with the  probability distribution of recognition types. Specifically, the accuracy for the $i$-th action type is $\left(1-p_i^o\right)\alpha(F^{\text{s}})$ with $i\ge 2$ since instances of the $i$-th type can be recognized correctly only after they are not detected as the static state by the action detection module. \revision{The accuracy for the static state is $(1-p^l) + p^l\alpha(F^{\text{s}})$ where $(1-p^l)$ and $p^l\alpha(F^{\text{s}})$ represent the probabilities that instances of the static state are correctly detected by the action detection module, and incorrectly detected by the action detection module but correctly detected by the CNN module, respectively.} Therefore, the average recognition accuracy of the action recognition task is given by 
\begin{equation}
	  A = \sum_{i=2}^{I}p_i^m\left(1-p_i^o\right)\alpha(F^{\text{s}}) + p_1^m \left((1-p^l) + p^l\alpha(F^{\text{s}})\right), \!\! \label{eq:acc}
\end{equation}
where $p_i^m = \dfrac{M_i}{\sum_{i=1}^{I}M_i}$ denotes the probability of the $i$-th type with $\sum_{i=1}^{I}p_i^m = 1$. 

\subsection{Sensing Delay}

\revision{The delay of the signal collection phase is constant, and the cost of the action detection module is small enough to be neglected since the main computation cost in the action detection module is the FFT operation. Therefore, we focus on the CNN module.\footnote{\revision{For example, according to test results, when $F^s$ is 200 Hz and the computation resource of the edge server is $2.64$ GHz, the computation delay of the action detection is around $2\times 10 ^{-5}$ s while that of the CNN module is around 0.75 s.}}} Recall that the input of the CNN module is matrix $\bm{G}$ and its size is $K\times (T^{\text{s}}F^{\text{s}})$. Meanwhile, the main part of the CNN is the convolutional layer, which causes the major computation delay of the CNN. According to \cite{cnn_time}, the computational complexity of convolutional layers is proportional to the input matrix size, i.e., $K\times (T^{\text{s}}F^{\text{s}})$. Thus, the overall computational complexity of the CNN module is $\mathcal{O} \left(K T^{\text{s}}F^{\text{s}}\right)$. \revision{Let $f^{\text{s}}$ denote the computation resource at the edge sever allocated to the CNN module.} The corresponding delay can be modeled as
\begin{equation}
	T^{\text{CNN}} = \frac{C^{\text{s}}K T^{\text{s}}F^{\text{s}}}{f^{\text{s}}}, \label{eq:cnn_delay}
\end{equation}
where $C^{\text{s}}$ is the computation cost of the CNN module per element.%\footnote{We assume that the edge server is equipped with a CPU for the inference of the CNN module. The GPU scenario will be studied in the future work. } 

Since we have added the action detection module, not every instance would pass through the CNN module. Thus, we utilize the average delay to represent the performance of sensing tasks. Based on the analysis in Section III-A, the probability of an instance passing through the CNN module is
\begin{equation}
	p^{\text{CNN}} = \sum_{i=2}^{I}p_i^m\left(1-p_i^o\right)  + p_1^m  p^l  , \label{eq:cnn_p}
\end{equation}
where the first part is for action types and the second part is for the static state. Combining the probability in (\ref{eq:cnn_p}) with the delay in (\ref{eq:cnn_delay}), the average delay of the sensing task is 
\begin{equation}
	T^{\text{s}} = \left( \sum_{i=2}^{I}p_i^m\left(1-p_i^o\right)  + p_1^m  p^l  \right)\frac{C^{\text{s}}K T^{\text{s}}F^{\text{s}}}{f^{\text{s}}}. \label{eq:delay}
\end{equation}

\subsection{Performance Comparison}

\revision{With the above analysis, we can identify the performance improvement of  the proposed sensing framework by comparing it with the conventional one where there is no action detection module.} The performance gain can be derived as shown in the following theorem.
\begin{thm} \label{thm:gain}
	Under the same recognition accuracy target and delay requirement, the proposed sensing framework needs less computation resource than the conventional scheme when the following condition is satisfied
	\begin{equation}
		1- \sum_{i=2}^{I} \frac{p_i^m \alpha(F^{\text{s}}) \sigma_{P,1} }{p_1^m \left(1- \alpha(F^{\text{s}})\right) \sigma_{P,i}} \exp\left( \frac{(\mu_{P,1} )^2}{2\sigma_{P,1}^2 }- \frac{(\mu_{P,i} )^2}{2 \sigma_{P,i}^2} \right)>0. \label{eq:gain_condition}
	\end{equation}
	The performance gain, defined as the ratio of the saved computation resource to the required computation resource in the conventional framework, is 
	\begin{equation}
		\rho  =  \min\left\{\frac{p_1^m(1-p^l)}{A}|_{A=\alpha(F^{\text{s}})},1-p^{\text{CNN}}|_{\eta =\eta_u} \right\}, \label{eq:gain}
	\end{equation}
where $\eta_u = \min\limits_{i\in \{2,\cdots, I\}}\mu_{P,i} $.
	\begin{IEEEproof}
		Please refer to Appendix \ref{proof:thm_gain}. 
	\end{IEEEproof}
\end{thm}
\begin{rem} \revision{
According to Theorem \ref{thm:gain}, we can verify the performance gain, that is, the proposed sensing framework can reduce the cost of computation resource without lowering the recognition accuracy.  In other words, compared with the conventional framework, it can improve the accuracy with the same computation resource. To understand (\ref{eq:gain_condition}), we can focus on the probability distribution of action, i.e., $\{p_i^m\}$, and the accuracy of the CNN module, i.e., $\alpha(F^{\text{s}})$. As the probability of the static state $p_1^m$ becomes higher, the condition is easier to be satisfied since the action detection module can avoid more unnecessary computation cost. Meanwhile,  as the accuracy of the CNN module increases, the condition is more difficult to satisfy, because more instances would be input into the CNN module to achieve higher average recognition accuracy. Based on the above analysis, the proposed sensing framework is suitable for smart home scenarios, since the sensing target (e.g., human) keeps in the static state most of the time. When the condition is met, there are two cases. The performance gain is $1-p^{\text{CNN}}|_{\eta = \eta_u}$ when the action detection module can almost perfectly distinguish between action instances and static instances, otherwise, it is $\dfrac{p_1^m(1-p^l)}{A}$. In the first case, the performance gain is only determined by the probability of passing through the CNN module, i.e., $p^{\text{CNN}}$. Furthermore, the relationship between them is negatively correlated since the computation cost increases with the probability of passing through the CNN module. In the second case, the performance gain is influenced by both the accuracy of the CNN module and the probability of the static state. The performance gain decreases because more instances are fed into the CNN module to achieve higher overall accuracy as the CNN accuracy increases. The performance gain increases with the probability of the static state. It is because, as the probability of the static state increases, the threshold in the action detection module can be set higher without pulling down the accuracy according to (21), and fewer instances are input into the CNN module. This eventually reduces the probability of computation.  Besides,  even in the worst case, the proposed sensing framework maintains the same performance as the conventional framework.}
\end{rem}
%In an extreme case, the condition (\ref{eq:gain_condition}) cannot be satisfied and there is no performance gain. Besides, the performance gain increases with the probability of the static state. It is because that as the probability of the static stateincreases, less instances are sent into the CNN module with the proposed action detection module, and thus  the probability of computation is reduced.

\section{Sensing Accuracy Maximization}
In this section, we formulate a mathematical problem to maximize the sensing accuracy and then decompose it into two subproblems for achieving its solution. 
\subsection{Problem Formulation}
\revision{Thus far, we have mathematically modeled the sensing performance including the average recognition accuracy in (\ref{eq:acc}) and the average sensing delay in (\ref{eq:delay}). In the following, we aim to solve the third challenge mentioned before, i.e.,  maximizing the sensing performance.} To be specific, we aim to maximize the sensing accuracy under the delay requirement, denoted by $T^{\text{s},\max}$, by optimizing the threshold and sampling rate. Meanwhile, we need to consider the delay requirement of each computation task since the computation and communication resource is limited. Let  $f^{\text{e}}$ denote the total computation resource at the edge server. Then, we formulate the sensing accuracy maximization problem, as
\begin{subequations} \label{pb_o}
\begin{eqnarray}
	\!\!\!&\max\limits_{\left\{  F^{\text{s}}, \eta, f^{\text{s}}, \tau^{\text{c}}_n, f_n  \right\}}	& A =  \sum_{i=2}^{I}p_i^m\left(1-p_i^o\right)\alpha(F^{\text{s}}) \nonumber \\ 
	&	&~~~~ + p_1^m \left((1-p^l) + p^l\alpha(F^{\text{s}})\right), ~~~~~ \label{pb_o:o}\\
	&\text{s.t.} & T^{\text{s}} \le T^{\text{s},\max},\label{pb_o:t_sen} \\
	&	& T_n \le T^{\max}_n,~\forall n, \label{pb_o:t_user}\\
	&	& \tau^{\text{s}}F^{\text{s}} + \sum_{n=1}^N \tau^{\text{c}}_n \le 1, \label{pb_o:comm} \\
	&	& f^{\text{s}} + \sum_{n=1}^N f_n \le f^{\text{e}},\label{pb_o:comp}\\
	&	& 0\le \eta \le \eta_u, \label{pb_o:eta}\\
	&	&  F^{\text{s}} \in \mathcal{Z}^+,   f^{\text{s}}, \tau^{\text{c}}_n, f_n \ge 0, \label{pb_o:0}
\end{eqnarray}
\end{subequations}
where $\mathcal{Z}^+$ denotes the set of positive integers. 
In the above, (\ref{pb_o:t_sen}) and (\ref{pb_o:t_user}) are the delay constraints for the sensing task and computation tasks, respectively, (\ref{pb_o:comm}) and (\ref{pb_o:comp}) are the communication and computation resource limitation constraints, respectively, and (\ref{pb_o:eta}) gives the range of the threshold  where the upper bound is $\eta_u$ that is the minimum power mean among all action types. The reason for setting an upper bound is that  the accuracy would drop rapidly if the threshold  is higher than $\eta_u$.

%\vspace*{-2ex}
\subsection{Problem Decomposition}
By jointly considering (\ref{pb_o:o}), (\ref{pb_o:t_sen}), and (\ref{pb_o:eta}), we can find that the effect of sampling rate on the sensing accuracy is quite complicated. \revision{Thus, we first give $F^{\text{s}}$ and then solve the accuracy maximization problem (\ref{pb_o}) with a given $F^{\text{s}}$. After that, by performing an exhaustive search on $F^{\text{s}}$, we can find the optimal solution to problem (\ref{pb_o}). To solve the accuracy maximization problem (\ref{pb_o}) with given $F^{\text{s}}$, we can find that the sensing and computation tasks compete for the computation resource. Meanwhile, we aim to maximize the sensing accuracy. Therefore, we can minimize the computation resource required for the computation tasks while satisfying their delay requirement, and allocate all  remaining resource to the sensing task. In this way, the accuracy maximization problem (\ref{pb_o}) for a given $F^{\text{s}}$ can be intuitively decoupled into two subproblems.} The first one is a resource allocation problem that minimizes the required computation resource by optimizing $\tau_n^{\text{c}}$ and $f_n$, as
\begin{eqnarray} \label{pb:resource}
	&\min\limits_{\left\{  f_n, \tau^{{\text{c}}}_n\right\} }	& \sum_{n=1}^N f_n, \\
	&\text{s.t.}	& \text{(\ref{pb_o:t_user}), (\ref{pb_o:comm}),  and (\ref{pb_o:0})}. \nonumber
\end{eqnarray}
The second one is a threshold selection problem that maximizes the sensing accuracy by optimizing the threshold with $ f^{\text{s}} = f^{\text{e}}- \sum_{n=1}^N f_n $, as
\begin{eqnarray} \label{pb:acc}
	&\min\limits_{   \eta   }	& A, \\
	&\text{s.t.}	& \text{(\ref{pb_o:t_sen}) and (\ref{pb_o:eta})}. \nonumber
\end{eqnarray}

In the following, we will solve the two subproblems in turn.

\subsection{Resource Allocation Strategy}
First of all, we can prove that problem (\ref{pb:resource})  is convex, as shown in Lemma 1.
\begin{lem} \label{lem:convex}
	Problem (\ref{pb:resource}) is a convex optimization problem.
\begin{IEEEproof}
	The objective function and constraints  (\ref{pb_o:comm}) and (\ref{pb_o:0}) are linear, and constraint (\ref{pb_o:t_user})  is a linear combination of two convex functions, i.e., $1/\tau^{\text{c}}_n$ and $1/f_n$. As a result, problem (\ref{pb:resource}) is convex, which ends the proof.
\end{IEEEproof}
\end{lem}

Based on Lemma \ref{lem:convex}, we utilize the Lagrangian method to solve problem (\ref{pb:resource}) and the corresponding partial Lagrangian function can be expressed as
\begin{align}
	\mathcal{L} =& \sum_{n=1}^N f_n + \sum_{n=1}^N\lambda_n \left(T_n - T^{\max}_n\right) \nonumber 
	\\ &+ \mu\left(\tau^{\text{s}}F^{\text{s}} + \sum_{n=1}^N \tau^{\text{c}}_n - 1\right)  ,
\end{align}
where $\lambda_n$ and $\mu$ are the Lagrange multipliers associated with the constraints (\ref{pb_o:t_user}) and (\ref{pb_o:comm}). Let $\left\{ f_n^{\star}, \tau_n^{{\text{c}},\star} \right\}$ denote the optimal solution to problem (\ref{pb:resource}). Then, by applying the Karush-Kuhn-Tucker (KKT) conditions, we can derive the following theorem.
\begin{thm} \label{thm:resource}
	The optimal solution to problem (\ref{pb:resource})  can be given by
	\begin{equation} 
		\left\{ 
		\begin{array}{l}
			\tau_n^{ {\text{c}},\star} =\dfrac{V_n}{T^{\max}_{n}}\left(\dfrac{1}{R_n}+\sqrt{\dfrac{C_n}{\mu^\star  R_n}} \right),\\
			f_n^\star =\dfrac{V_n}{T^{\max}_{n}}\left( C_n +\sqrt{\dfrac{\mu^\star C_n  }{ R_n}} \right) ,
		\end{array}
		   \right.\! \!\!n = 1,2,\cdots, N,
	\end{equation}
where $\mu^{\star}$ is the optimal Lagrange multiplier, as
\begin{equation}
	\mu^{\star}=\frac{\sum_{n=1}^N \frac{V_n}{T_n^{\max}} \sqrt{\frac{C_n}{R_n}} }{\left(1-\tau^{\text{s}}F^{\text{s}}-\sum_{n=1}^N\frac{V_n}{T_n^{\max}R_n} \right) }.
\end{equation}
\begin{IEEEproof}
	Please refer to Appendix \ref{proof:thm_convex}.
\end{IEEEproof}
\end{thm}

\begin{rem}
Theorem \ref{thm:resource} gives the optimal resource allocation strategy. The allocated amount of resource to each computation task is negatively related to the data rate and is positively related to the computation intensity and task size. The computation resource decreases sublinearly with the data rate  exponentially by 1/2. Moreover, we should note that the remaining computation resource is all allocated to the sensing task. Thus, the computation resource for the sensing task increases sublinearly with exponential 1/2 of the data rate.
\end{rem}

\subsection{Threshold Selection Policy}

Now, we focus on problem (\ref{pb:acc}). To solve it, we need to explore the properties of the objective function and constraint (\ref{pb_o:t_sen}), i.e., $A$ and $T^{\text{s}}$. $T^{\text{s}}$ decreases with  $\eta\in \left[0,  \eta_u \right]$ since $p_i^o$ monotonically increases with   $\eta $ and $p^l$ monotonically decreases  with  $\eta $. Thus, delay constraint (\ref{pb_o:t_sen}) is equivalent to $\eta \le \eta^{T}$, where $\eta^{T}$ satisfies that  $T^{\text{s}} = T^{\text{s,max}}$ and can be obtained via the exhaustive search algorithm. The  computational complexity is $\mathcal{O}\left(  \log(\eta_u/\epsilon)\right)$, where $\epsilon$ is the tolerance of accuracy. 

As for $A$, it is a concave function with $\eta \in [\mu_{P,1},\eta_u]$ since both $p^l$ and $p^o_i$  are convex with $\eta \in [\mu_{P,1},\eta_u]$. \revision{However, when $\eta \in [0,\mu_{P,1}]$, $A$ is not concave since $p^l$ is not convex, and thus we cannot obtain the optimal solution directly. To address it, we adopt piecewise linear approximation (PLA) for $p^l$. First of all, the set $[0,\mu_{P,1}]$ is divided into $M$ subsets and the $m$-th subset is $\left[\dfrac{(m-1)\mu_{P,1}}{M},\dfrac{m\mu_{P,1}}{M}\right]$. In each subset, $p^l$ is approximated as a linear function. Then, $A$ is a concave function with $\eta$ belonging to each subset since $p^l$ is linear and $p^o_i$ is convex. Let $\eta^{A,m}$ denote the threshold that maximizes $A$ within the $m$-th subset and it can be obtained via the Golden-section search algorithm \cite{search} with the computational complexity of $\mathcal{O}\left(  \log(\mu_{P,1}/M/\epsilon)\right)$.} Besides, the threshold that maximizes $A$ within $[\mu_{P,1},\eta_u]$ is denoted by $\eta^{A,M+1}$. Then, the threshold that  maximizes $A$ within $[0,\eta_u]$, denoted by $\eta^{A}$, is given by 
\begin{equation}
	\eta^{A} = \max_{m \in \left\{ 1,\cdots,M+1\right\} } \eta^{A,m}.
\end{equation}
The computational complexity for calculating $\eta^{A}$ is $\mathcal{O}\left( M \log(\mu_{P,1}/M/\epsilon) + \log((\eta_u-\mu_{P,1})/\epsilon) +M \right) $.

By jointly considering the objective function and delay constraint, there are two cases. 
\begin{itemize}
	\item When $\eta^{A} \ge \eta^T$,  $\eta^{A} $ is a feasible solution to problem  (\ref{pb:acc}) and the maximum accuracy is achieved when $\eta = \eta^{A} $.
 	\item When $\eta^{A} < \eta^T$,  the delay requirement cannot be satisfied when $\eta = \eta^{A} $ and the accuracy decreases with $\eta>\eta^{A}$. Thus, the maximum accuracy is achieved when $\eta = \eta^T$.
\end{itemize}
Combining the above two cases, the optimal solution to problem (\ref{pb:acc}) is shown in the following theorem.
\begin{thm} \label{thm:eta}
	The optimal threshold selection policy is given by 
	\begin{equation}
		\eta^\star = \max\left(\eta^T, \eta^A\right).
	\end{equation}
\end{thm}

\begin{comment}
\begin{figure}[h]
	\vspace{-2ex}
	\centering
	\includegraphics[width=0.5\linewidth]{fig/eta.pdf}
	\vspace{-2ex}
	\caption{The threshold selection policy}
	\label{fig:eta}
\end{figure}
\end{comment}
\begin{rem}
	From Theorem  \ref{thm:eta}, the delay requirement may limit the sensing accuracy. We should note that when the accuracy is limited by the delay constraint in the proposed sensing framework, there is no feasible  solution in the conventional framework. \revision{This is because the proposed sensing framework is equivalent to the conventional one when $\eta=0$, and  $\eta^T$ should be set higher than 0 to meet the delay constraint as $T^{\text{s}}$ decreases with  $\eta\in \left[0,  \eta_u \right]$.} Furthermore, the accuracy of the  proposed framework is higher than that of the conventional one when the accuracy is not limited by the delay constraint. As a result, the proposed framework performs better than the conventional one. 
\end{rem}

\subsection{Overall Algorithm}

{\begin{algorithm}[t]
		\caption{Overall Algorithm for Solving Problem ({\ref{pb_o}}).}
		\label{alg:overall}
		\DontPrintSemicolon
		Initialize the maximal error tolerance $\epsilon>0$ and the maximal accuracy $A^{\star}=0$;\;
		\For{$F^{ \mathrm{ \text{s}}} \in \left\{1,\cdots,F^{\text{s}}_u\right\}$}{
			\% Resource allocation strategy;\;
			Obtain the optimal resource allocation strategy $\{\tau_n^{{\text{c}},\star},f_n^{\star}\}$ with Theorem \ref{thm:resource};\;
			\% Threshold Selection Policy;\;
			Obtain $\eta^\star$ with Theorem \ref{thm:eta};\;
			Calculate the overall accuracy $A$;\;
			\If{$A^{\star} < A $}{$A^{\star} = A$;}
		}
	\end{algorithm}
}

Before presenting the overall algorithm, we give an upper bound of $F^{\text{s}}$ in the following. 
\begin{lem}\label{lem:upper}
	An upper bound of $F^{\text{s}}$ is given by
	\begin{equation}
	  	F^{\text{s}}_u =\left\lfloor \frac{1}{\tau^s}\!\left(1 - \frac{ N \min\limits_n V_n f^{\text{e}} /\min\limits_n R_n }{    \max\limits_n T^{\max}_n f^{\text{e}} -N \min\limits_n V_n \min\limits_n  C_n }\right) \right\rfloor.\!\! \label{eq:fs_up}
	\end{equation}
\begin{IEEEproof}
	Please refer to Appendix \ref{proof:lem_up}.
\end{IEEEproof}
\end{lem}

Based on the above analysis, we can obtain  the overall algorithm to problem (\ref{pb_o}), as shown in Algorithm \ref{alg:overall}. Specifically, we perform exhaustive searching on $F^{\text{s}}$. In each loop iteration, we solve problems (\ref{pb:resource}) and (\ref{pb:acc}) with Theorems \ref{thm:resource} and \ref{thm:eta}, respectively. By comparing the accuracy under different $F^{\text{s}}$, we can find the maximal accuracy of problem (\ref{pb_o}), denoted by $A^{\star}$. The computational complexity of each loop iteration is $\mathcal{O}\left(N+M \log(\mu_{P,1}/M/\epsilon) + \log((\eta_u-\mu_{P,1})/\epsilon) +M\right)$, where the computational complexity for obtaining the optimal resource allocation strategy is $\mathcal{O}\left(N\right)$. Since the number of loop iterations is $F^{\text{s}}_u$, the total computational complexity of Algorithm \ref{alg:overall} is 
\begin{equation}
	\mathcal{O}\left(F^{\text{s}}_u\left(N+M \log(\mu_{P,1}/M/\epsilon) + \log(\eta_u/\epsilon) +M\right)\right).
\end{equation}
\revision{Furthermore, Algorithm 1 is based on the exhaustive search method. Meanwhile, each loop solves the resource allocation problem (28) with Theorem 2 and the threshold selection problem (29) with Theorem 3. Thus, the convergence of Algorithm 1 can be  guaranteed.}

\revision{To tackle extreme scenarios where the computation ability of the BS is limited, we also introduce a low-complexity algorithm. Specifically, in each loop, we can set $M=1$ for reducing the computational complexity of solving the threshold selection policy optimization problem to $\mathcal{O}\left( \log(\mu_{P,1}/\epsilon) + \log(\eta_u/\epsilon) \right)$. Regarding the number of loops, we observe the sensing accuracy generally increases with the sampling rate according to Fig. \ref{fig:acc_Fs} in Section V-C. Thus, we heuristically seek to find the maximum sampling rate that ensures there is a feasible solution to problem (27) with binary search algorithm. As a result, the computational complexity of the low-complexity algorithm is $\mathcal{O}\left(\log(F^{\text{s}}_u) \left(N+\log(\mu_{P,1}/\epsilon) + \log(\eta_u/\epsilon) \right)\right)$. }

\section{Test Results}
In this section, we conduct experiments to validate the effectiveness of the proposed algorithm.

\subsection{Test Parameters} \label{sec:sim_para}

\begin{figure}[t]
	\vspace{-2ex}
	\centering
    \subfigure[Real-world experiment scenario.]{
		\centering
		\includegraphics[width=0.8\linewidth]{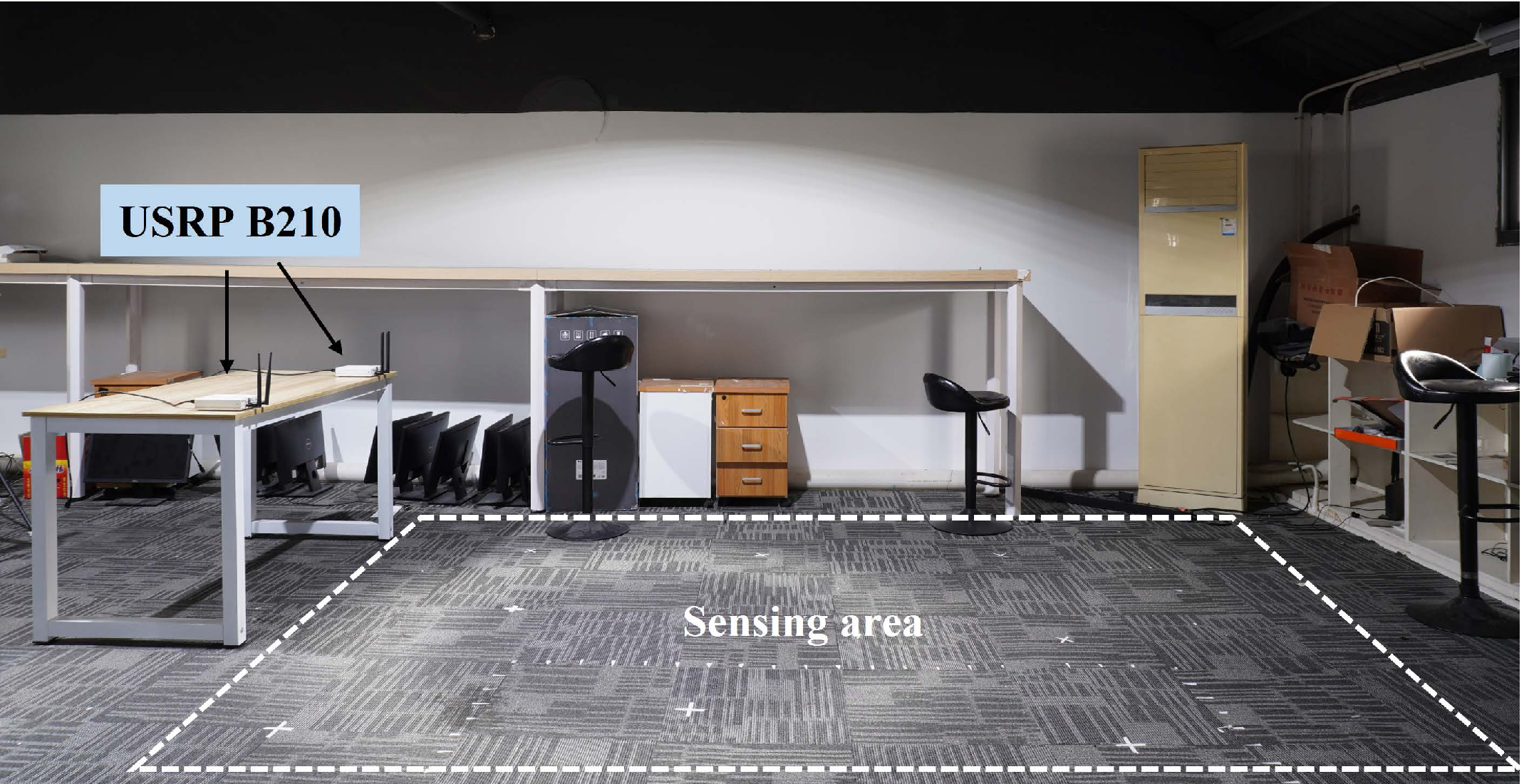}
	}
    \subfigure[Schematic diagram of the experiment scenario.]{
		\centering
		\includegraphics[width=0.85\linewidth]{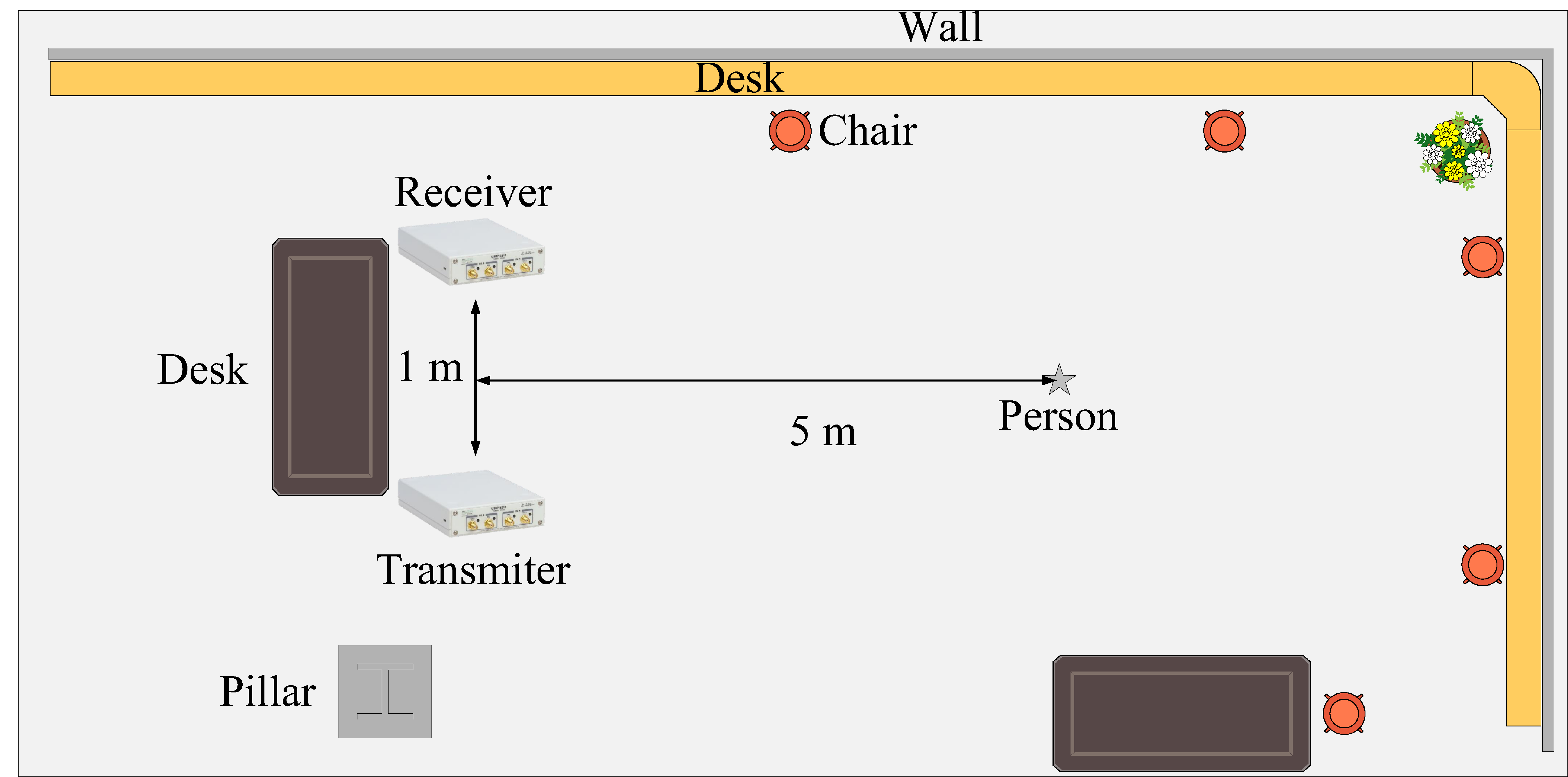}
	}
	\vspace{-2ex}
	\caption{\revision{The experiment setup for the sensing task.}}
	\label{fig:exp_setup}
\end{figure}

The test settings are provided as follows unless otherwise specified. We consider a  triple-functional BS with a radius of 300 m, and 15 devices are randomly located in the coverage. \revision{According to the long term evolution (LTE) standards \cite{LTE}, the system bandwidth is 4 MHz and the noise spectral density is -174 dBm/Hz.  The channel gain between each device and the BS is generated by following the path loss model: $128.1+37.6\log_{10}\left(d[km]\right)$, where $d$ denotes the distance between each device and the BS  in kilometer and  the small-scale fading is set to Rayleigh distributed with uniform variance.} The transmit power of each device is set as $24$ dBm. The total computation resource at the edge server is set as $40$ GHz. \revision{For the computation task at the device, the data size and computation intensity follow the uniform distribution with $V_n  \in [0.3, 1]$ Mbits and $C_n  \in [400,1000]$ CPU cycles/s, respectively, and the delay tolerance of all computation tasks is 0.4 s \cite{MEC_model}.}

For the sensing task, we consider an OFDM radar signal, where the FFT number is 64 and the cyclic prefix length is 16. The time of one sensing signal transmission slot is 128 us. The experiment setup is shown in Fig. \ref{fig:exp_setup}. Specifically, we use one USRP B210 to generate OFDM radar signals and one USRP B210 to collect radar echoes. Eight volunteers are invited to perform eight different behaviors, i.e., standing still, kicking, raising a hand, waving, bending down,  walking, sitting down, and standing up, where standing still refers to the static state. We totally collect 3,200 CSI matrices as a dataset, where 2,560 CSI matrices are used for training and 640 CSI matrices are used for testing. We adopt ResNet50 for the CNN module. It is trained on a Linux server equipped with four NVIDIA GeForce GTX 3080 GPUs and tested on a Linux personal computer (PC) equipped with an Intel i7-8700K CPU, which is regarded as the edge server.  

The detailed procedures of the simulation and experiment are presented in Fig. 7. Based on the sensing setting, we first collect the sensing dataset with USRP B210. The dataset is used for fitting $\lambda_i$, $r_i$, and $\sigma_{d,i}^2$ mentioned in Proposition 1 and training the CNN module to obtain the accuracy function $\alpha(F^s)$. The results can verify the proposed sensing models. Next, under the communication setting,  the performance of the proposed Algorithm 1 can be confirmed with  $\lambda_i$, $r_i$, $\sigma_{d,i}^2$, and $\alpha(F^s)$. \revisionv{Note that when there is no feasible solution, the accuracy is set as $12.5\%$ which is a lower bound of the action recognition task and is achieved with the random guess. }

\begin{figure}[t]
	\vspace{-2ex}
	\centering
	\includegraphics[width=1\linewidth]{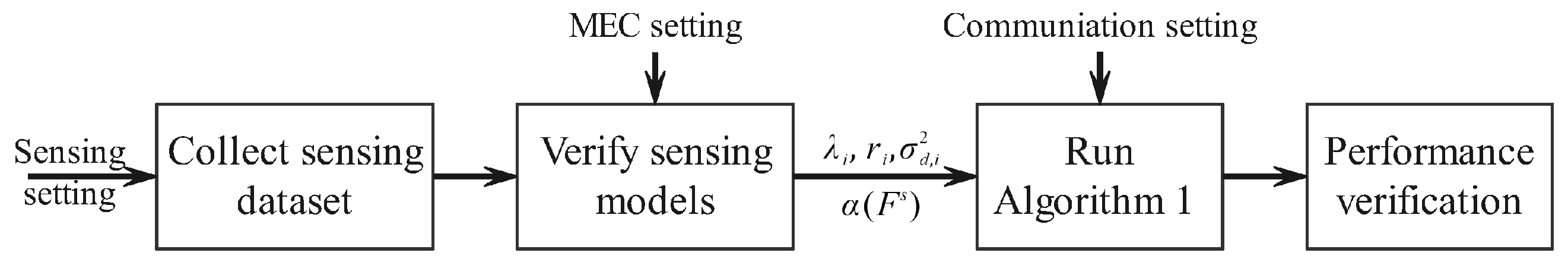}
	\vspace{-2ex}
	\caption{\revision{The detailed procedures of the simulation and experiment.}}
	\label{fig:sim_flow}
	\vspace{-3ex}
\end{figure}

\subsection{Model Verification}

\begin{figure*}[t]
	%\vspace{-2ex}
	\centering
	\begin{minipage} {0.64 \linewidth}
		\vspace{-5ex}
		%\centering
	\subfigure[The fitted curves of the power mean \protect\\  for eight types.]{
		\centering
		\includegraphics[width=0.5 \linewidth]{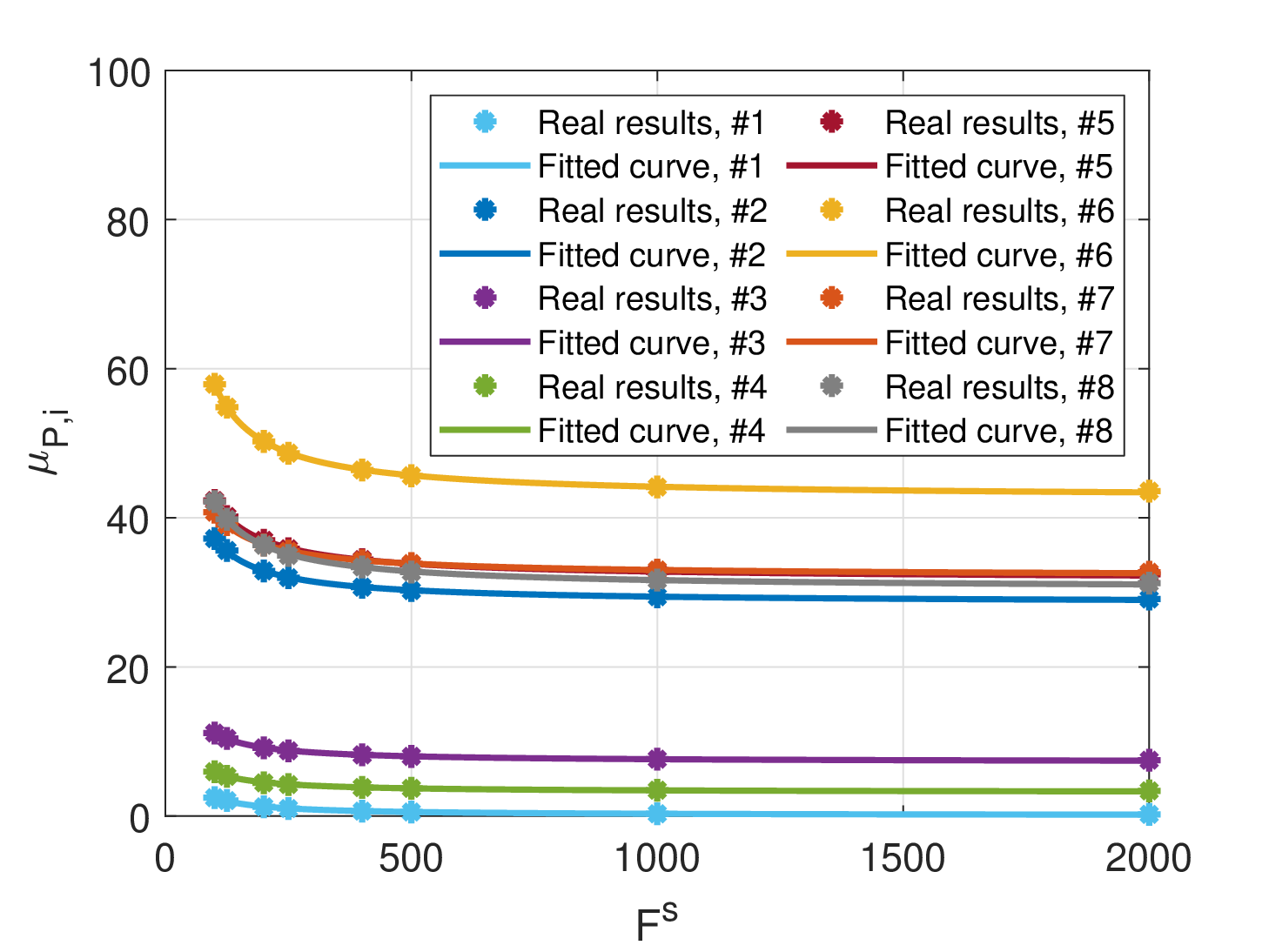}
	}
	\subfigure[The fitted curves of the power variance \protect\\  for eight types.]{
		\centering
		\includegraphics[width=0.5  \linewidth]{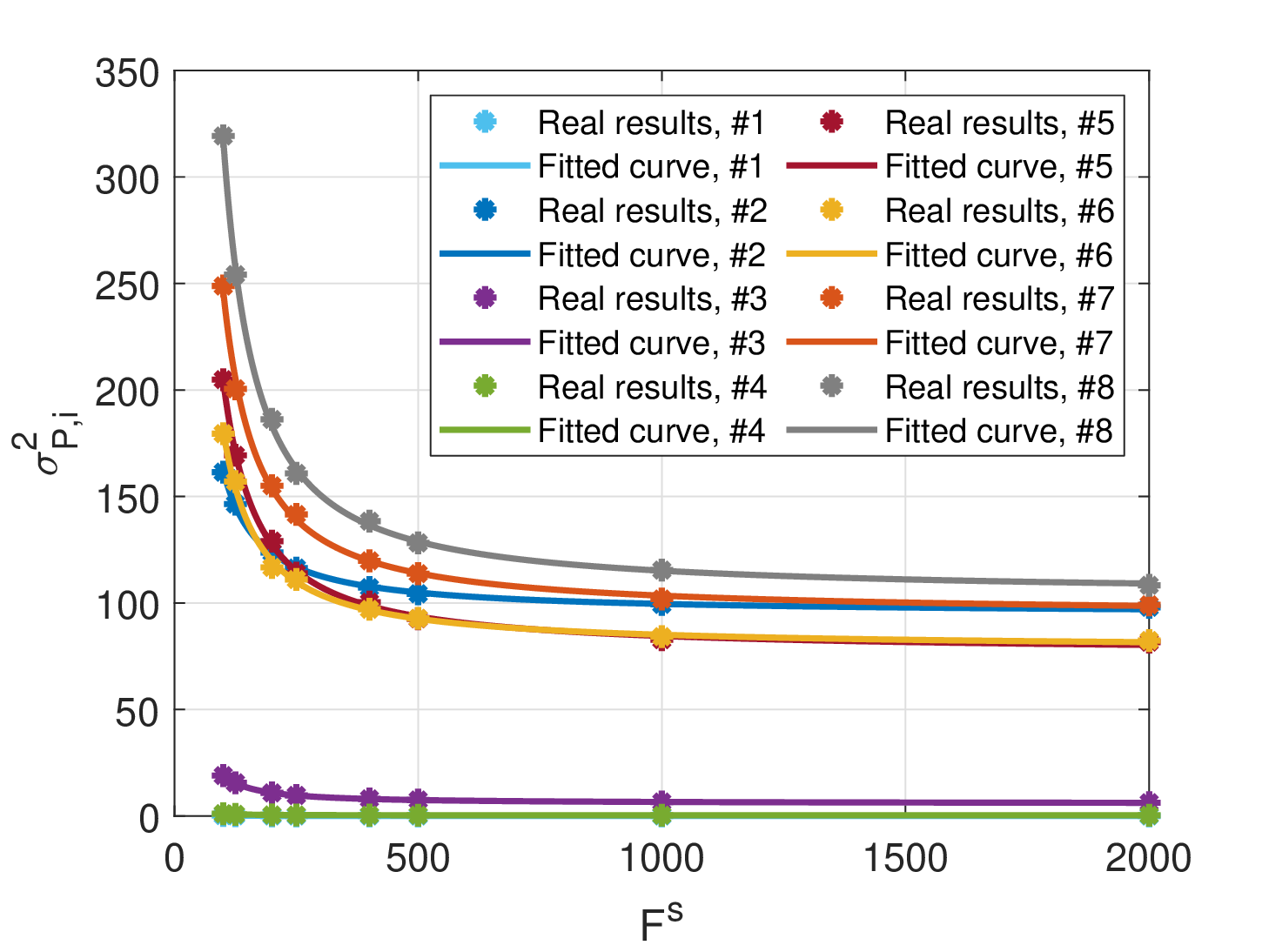}
	}
	\vspace{-2ex}
	\caption{The verification of the mathematical model for the power of high-frequency components. }
	\label{fig:mu_sigma}
	\end{minipage}
	\begin{minipage} {0.32 \linewidth}
		\centering
		\includegraphics[width=1\linewidth]{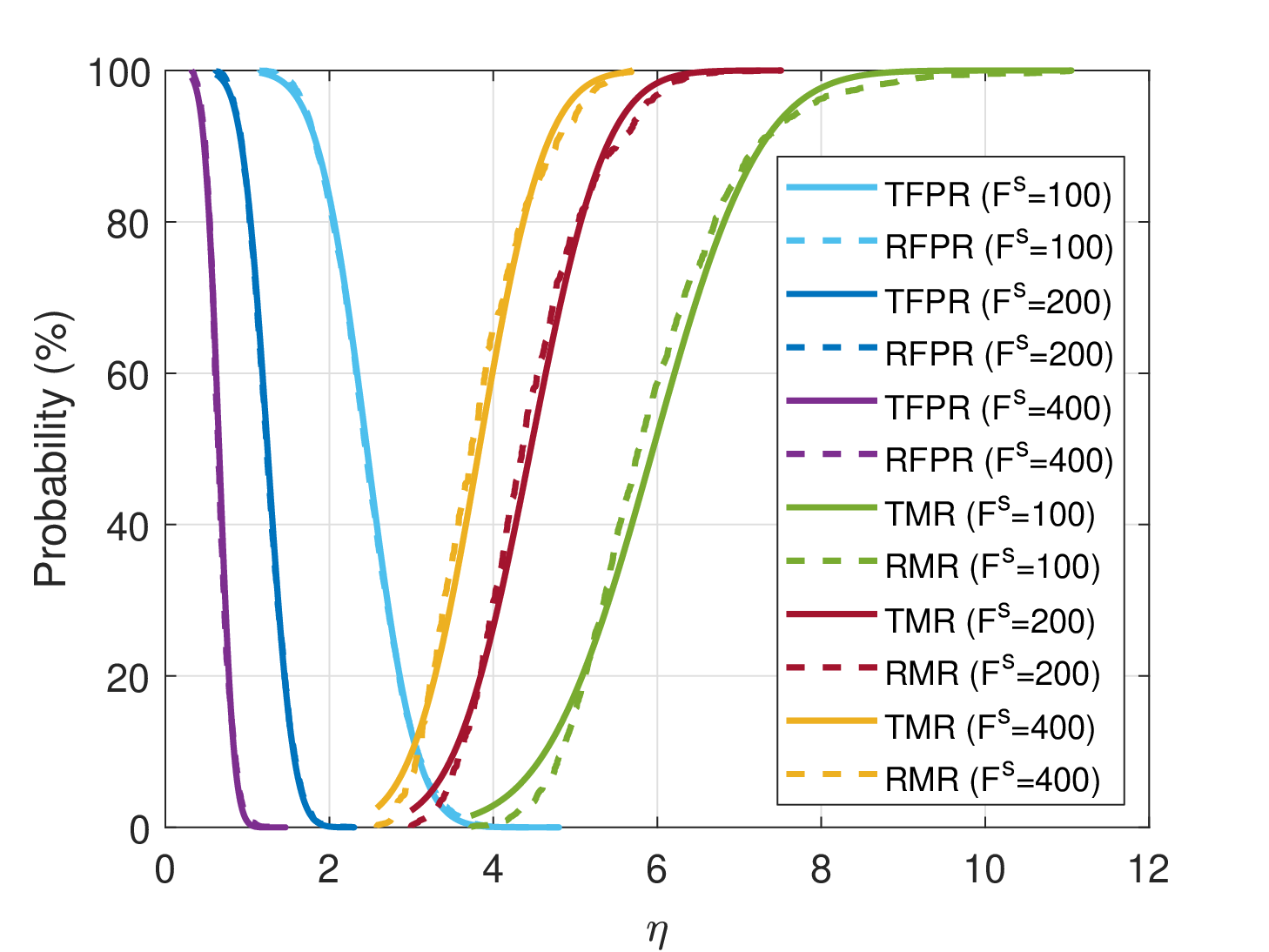}
		\vspace{-5ex}
		\caption{\revision{The verification of miss rate and false positive rate models in Proposition 1. ``TFPR'' represents the theoretical false positive rate. ``RFPR'' represents the real-world false positive rate. ``TMR'' represents the theoretical miss rate. ``RMR''  represents the real-world miss rate.}}
		\label{fig:verification_p}
	\end{minipage}
	\vspace{-4ex}
\end{figure*}

\begin{figure*}[t]
	\centering
	\begin{minipage} {0.32 \linewidth}
	\includegraphics[width=1 \linewidth]{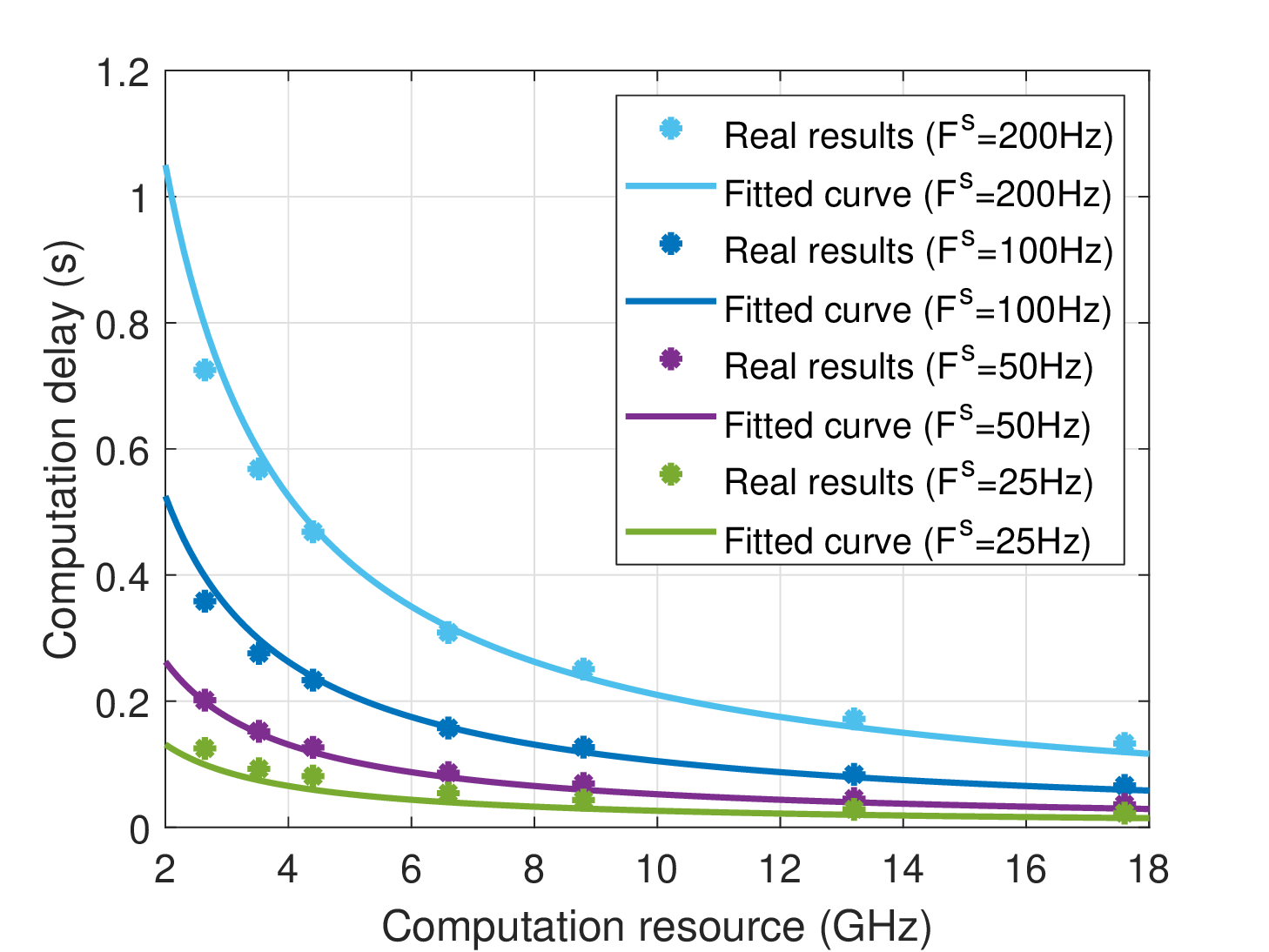}
	\vspace{-4ex}
	\caption{{The real-world computation delay of  \protect\\ the CNN module with different sampling rates.}}
	\label{fig:delay_cnn}
	\end{minipage}
	\begin{minipage} {0.64\linewidth}
		\subfigure[Accuracy vs. number of iterations/loops.]{
			\centering
			\includegraphics[width=0.5 \linewidth]{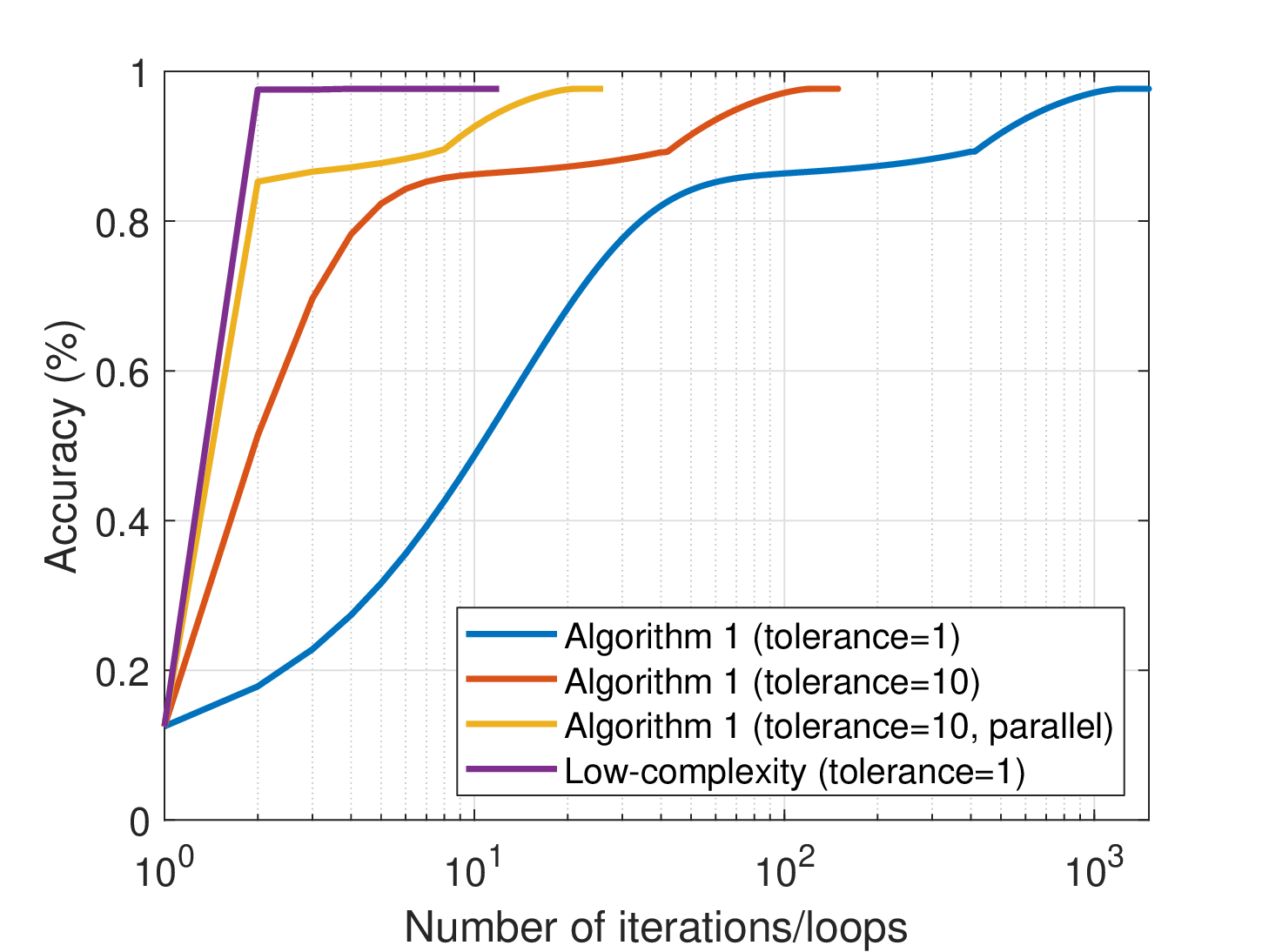}
		}
		\subfigure[Residual vs. number of iterations/loops]{
			\centering
			\includegraphics[width=0.5 \linewidth]{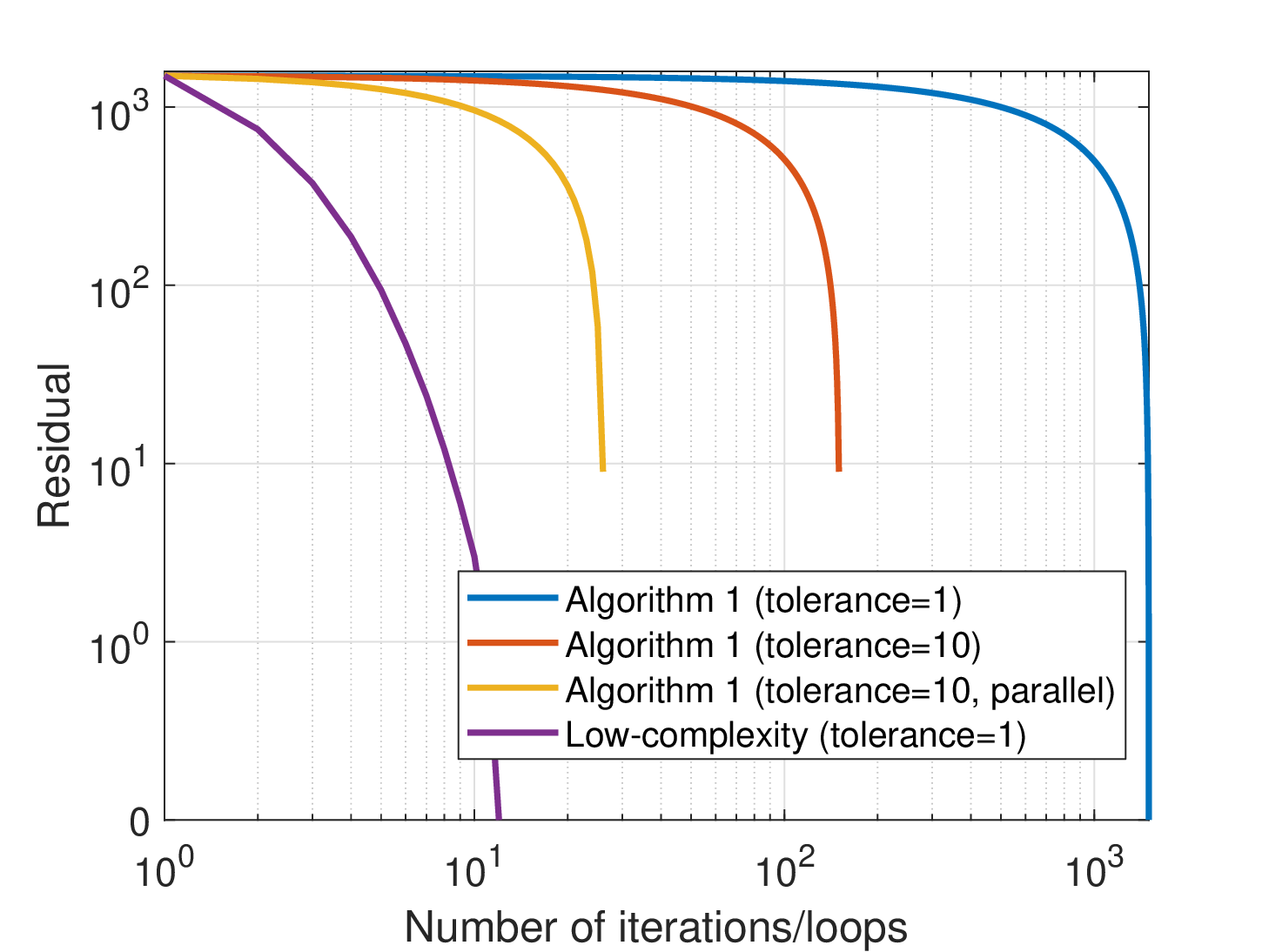}
		}
		\vspace{-3ex}
		\caption{Convergence behavior of the proposed algorithms.}
		\label{fig:converage}
	\end{minipage}	
	\vspace{-3ex}
\end{figure*}

In Section III, we have proposed two mathematical models: one is a power model of high-frequency components, i.e., $P$, for describing the miss rate and false positive rate in the action detection module, and the other is a delay model for the CNN module. Here, we use the collected real-world sensing dataset to verify both models. First of all, Fig. \ref{fig:mu_sigma} shows the mean and variance of $P$ for eight types and the corresponding fitted curves, respectively. The real mean and variance of each type are calculated based on the collected CSI matrices, and the corresponding fitted curves are obtained by following (\ref{eq:mu}) and (\ref{eq:sigma}), respectively. It can be observed that both the mean and variance decrease with the sampling rate and the fitted curves well match the real results under different sampling rates. \revision{Furthermore, we also plot Fig. \ref{fig:verification_p} to verify Proposition 1 and describe the relationship among miss rate, false positive rate, sampling rate, and threshold.   We only plot the miss rate of one action type due to the page limit. The remaining action types show similar results.  It can be observed that the real-world results match the theoretical results well, which further proves the effectiveness of our proposed model for the action detection module.   }

In Fig. \ref{fig:delay_cnn}, we show the real-world computation delay of the CNN module with the Linux PC and the corresponding fitted curves based on (\ref{eq:cnn_delay}). It can be seen that the computation delay increases with the sampling rate but decreases with the allocated computation resource. Moreover, the fitted curves match the real delay well. This verifies the effectiveness of the proposed delay model for the CNN module.

\subsection{Algorithm Investigation and Performance Comparison}

\revisionv{First of all, we show the convergence behavior of  Algorithm 1 and the low-complexity algorithm in Fig. \ref{fig:converage}. Since Algorithm 1 is based on the exhaustive search algorithm, it requires more than 1,000 iteration loops when the error tolerance (i.e., the step between two adjacent sampling rates) is set as 1. To accelerate convergence, the tolerance can be set to 10, and the number of loops is about 100. The final accuracy with the error tolerance being 10 is almost the same as that with the error tolerance being 1. Moreover, Algorithm 1 can be performed in parallel. Specifically, in each loop of Algorithm 1, the overall accuracy of different sampling rates can be calculated in parallel. The edge server used in our test contains 6 CPU cores and the number of loops can be reduced to about 30. As for the low-complexity algorithm, it is based on the binary search algorithm and the number of iterations is about 10, which verifies its high efficiency. Besides, the final accuracy of the low-complexity algorithm is almost the same as that of Algorithm 1, which verifies the high performance of the low-complexity algorithm.}

\begin{figure*}[t]
	%\vspace{-3ex}
	\centering
	\begin{minipage}[t]{0.32\linewidth}
		%\vspace{-2ex}
		\centering
		\includegraphics[width=1 \linewidth]{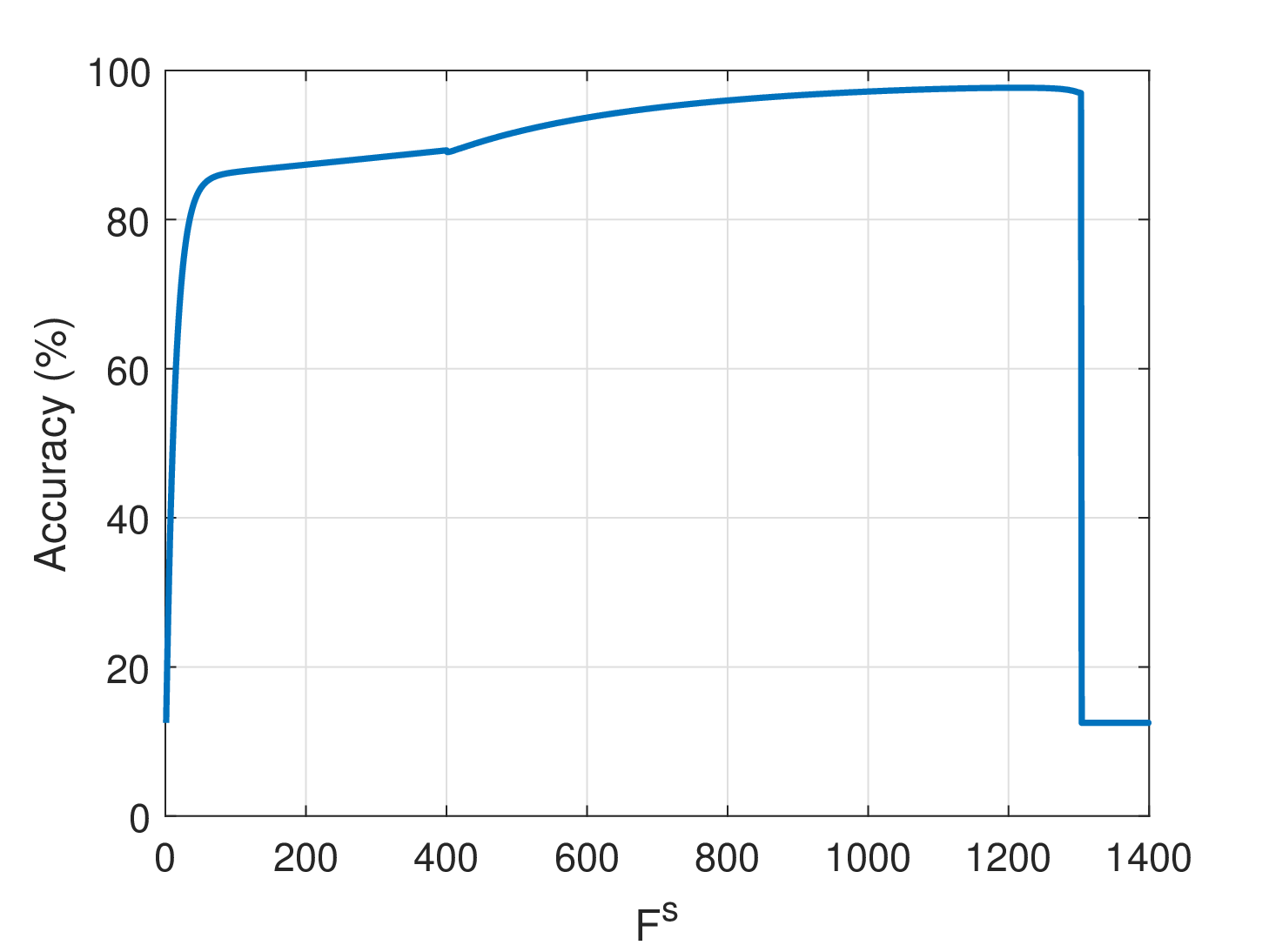}
		\vspace{-5ex}
		\caption{\revision{The optimized accuracy under the given sampling rate.}}
		\label{fig:acc_Fs}
	\end{minipage}
	\begin{minipage}[t]{0.32 \linewidth}
		%\vspace{-2ex}
		\centering
		\includegraphics[width=1 \linewidth]{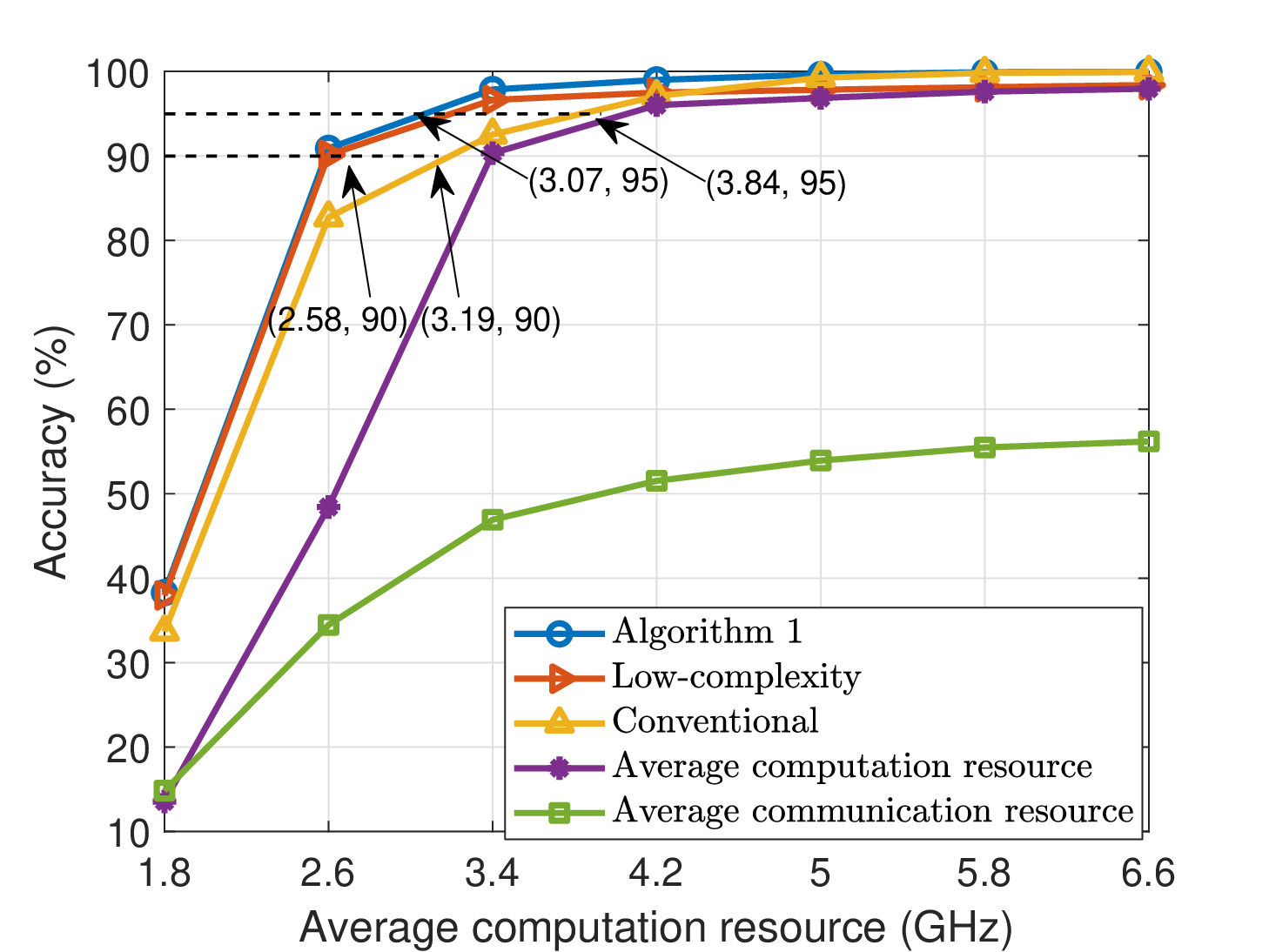}
		\vspace{-5ex}
		\caption{\revision{Average computation resource vs. accuracy.}}
		\label{fig:fcomp}
	\end{minipage}
		\begin{minipage}[t]{0.32 \linewidth}
		%\vspace{-2ex}
		\centering
		\includegraphics[width=1 \linewidth]{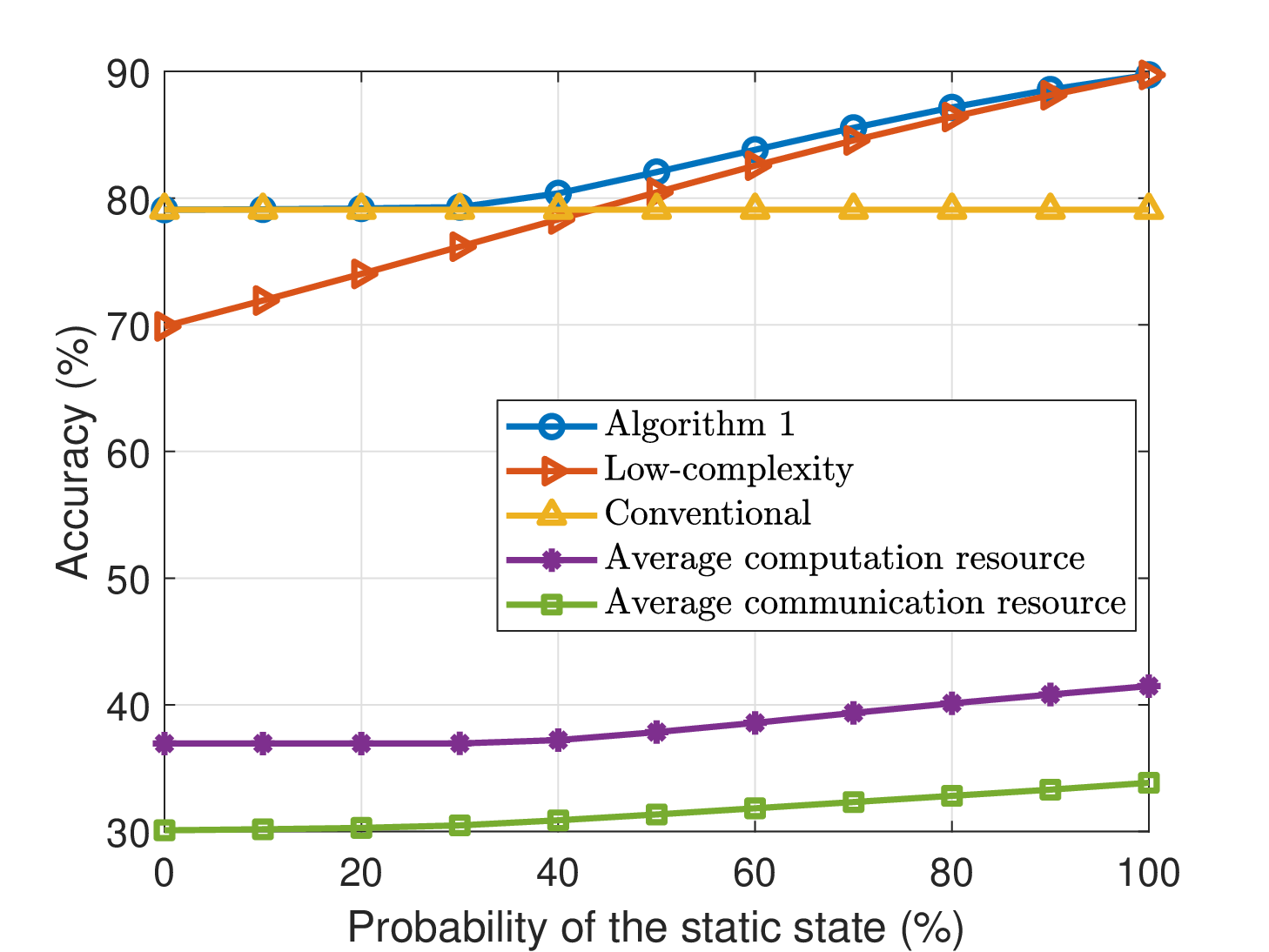}
		\vspace{-5ex}
		\caption{\revision{Probability of the static state vs. accuracy.}}
		\label{fig:pm}
	\end{minipage}
\vspace{-3ex}
\end{figure*}

Next, we also plot the optimized sensing accuracy under the given sampling rate, as shown in Fig. \ref{fig:acc_Fs}.  From the figure,  the optimized sensing accuracy substantially increases with the sampling rate. \revisionv{However, when the sampling rate is too high, there is no feasible solution and the sensing accuracy is set as 12.5\% as we have mentioned before.} This is because that the delay requirement of the sensing task cannot be satisfied when the sampling rate is too high.

\begin{figure*}[t]
	%\vspace{-3ex}
	\centering
	\begin{minipage}[t]{0.32 \linewidth}
		%\vspace{-2ex}
		\centering
		\includegraphics[width=1 \linewidth]{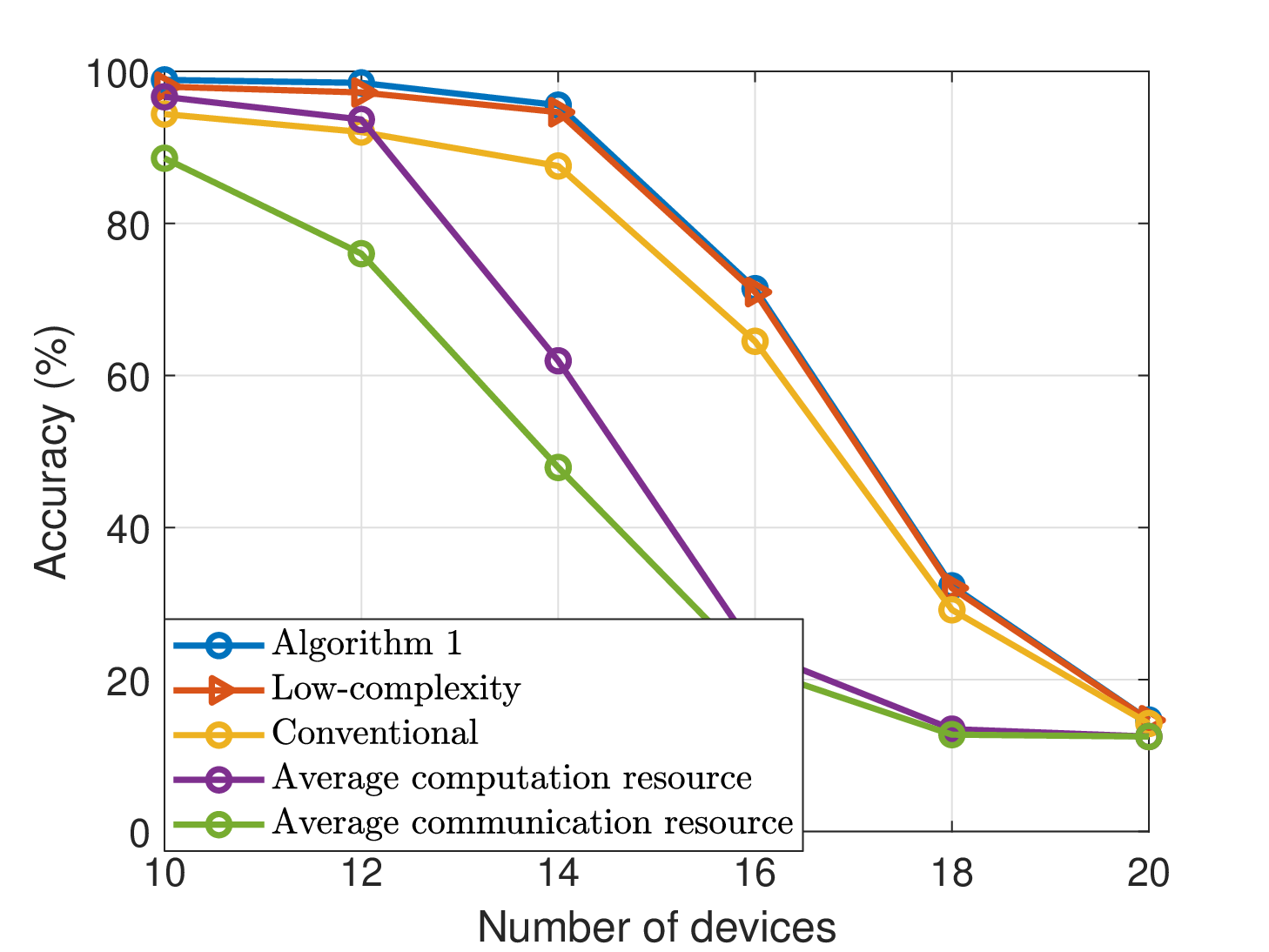}
		\vspace{-5ex}
		\caption{\revision{The number of devices vs. accuracy.}}
		\label{fig:nuser}
	\end{minipage}
		\begin{minipage}[t]{0.32 \linewidth}
		%\vspace{-2ex}
		\centering
		\includegraphics[width=1\linewidth]{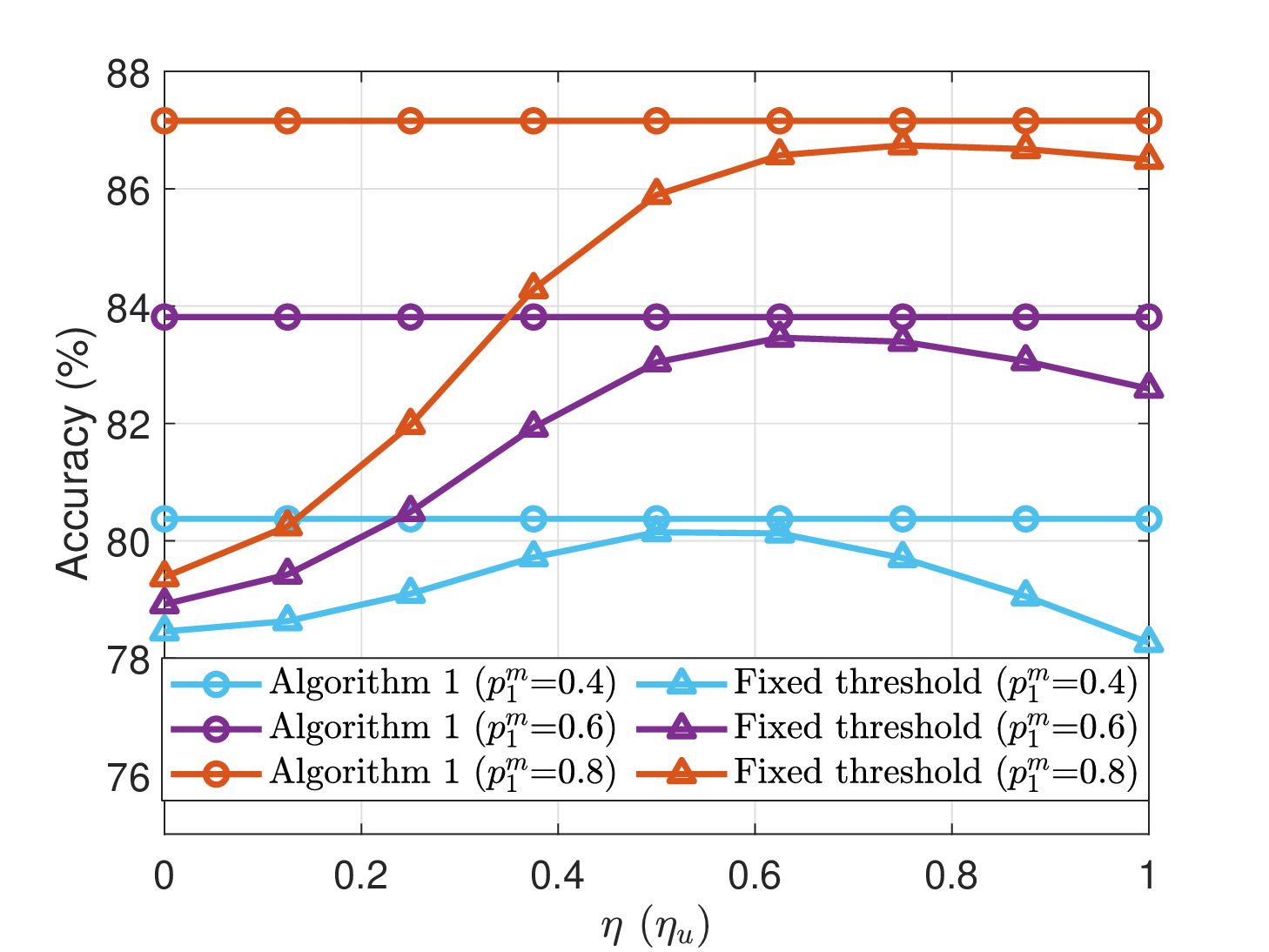}
		\vspace{-5ex}
		\caption{\revision{The influence of the threshold.}}
		\label{fig:s_eta}
	\end{minipage}
	\begin{minipage}[t]{0.32 \linewidth}
		%\vspace{-2ex}
		\centering
		\includegraphics[width=1 \linewidth]{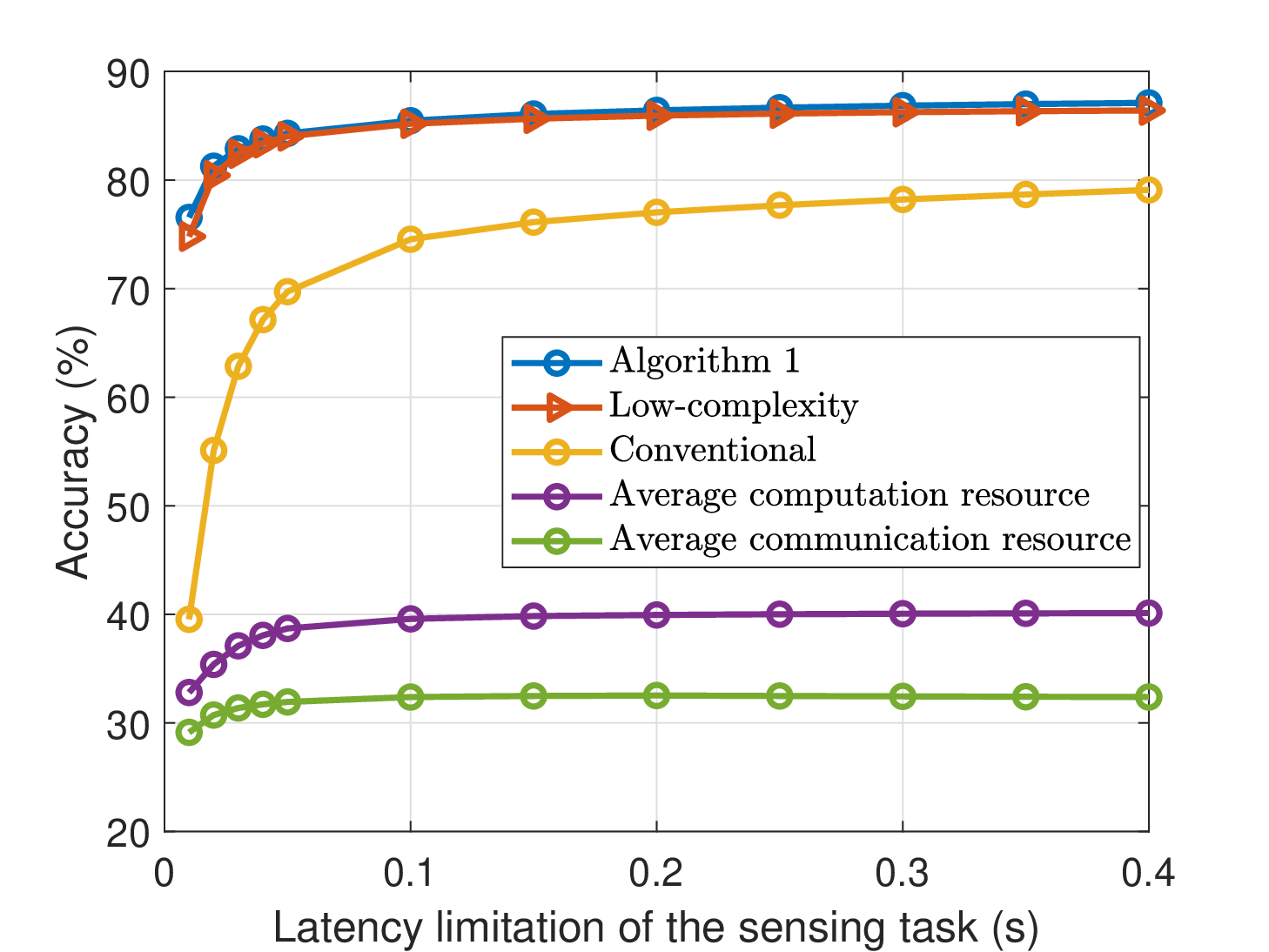}
		\vspace{-5ex}
		\caption{\revision{Delay requirement of sensing task vs. accuracy.}}
		\label{fig:tmax}
	\end{minipage}
	\vspace{-3ex}
\end{figure*}

To show the performance advantage of the sensing framework and the proposed algorithm, we select the following benchmark algorithms. 
\begin{itemize}
	\item \revision{Conventional scheme. The action detection module is not adopted in this scheme and the CSI matrix is directly input into the CNN module \cite{ISAC_learning}.} The resource allocation strategy is similar to the proposed algorithm. To be specific, it can be obtained by setting the threshold as 0 in Algorithm \ref{alg:overall}.
	\item Average computation resource scheme. The computation resource is equally allocated to each device and sensing task, i.e., $f_n = \dfrac{f^{\text{e}}}{N+1},\forall n$ and $f^{\text{s}} =  \dfrac{f^{\text{e}}}{N+1}$. The communication resource allocation strategy and the threshold selection policy are the same as those in the proposed algorithm. 
 	\item Average communication resource scheme. The communication resource is equally allocated to each device and sensing task, i.e., $\tau_n^{\text{c}} = \dfrac{1}{N+1},\forall n$ and $F^{\text{s}} =  \left\lfloor \dfrac{1}{(N+1)\tau^{\text{s}}} \right\rfloor$. The computation resource allocation strategy and the threshold selection policy are the same as those in the proposed algorithm. 
\end{itemize}

Fig. \ref{fig:fcomp} shows the effect of the average computation resource at the edge server on the accuracy. From the figure, we can observe that our proposed scheme provides the best performance among four schemes. Specifically, when the overall accuracy is $90\%$, our proposed scheme requires $2.58$ GHz average computation resource at the edge server, while the conventional scheme requires $3.19$ GHz average computation resource. The proposed sensing framework saves about $19.1\%$ computation resource, and this gap becomes more remarkable when accuracy is $95\%$.  This verifies the effectiveness of the proposed sensing framework.  When the computation resource is sufficient, the accuracy gap between the proposed scheme and the conventional scheme becomes very small. In this case, the accuracy of the CNN module is almost $100\%$ and all instances are input into the CNN module. Moreover, compared with the average computation resource scheme, our proposed scheme achieves higher performance when the computation resource is insufficient and shows the similar performance when the computation resource is sufficient. It demonstrates the effectiveness of the proposed resource allocation strategy. {Besides, the average communication resource scheme achieves much lower accuracy than the proposed scheme even if the computation resource is sufficient since its sampling rate is given and thus the accuracy is limited. \revision{Furthermore, the performance of the proposed low-complexity algorithm is close to that of Algorithm \ref{alg:overall}, which confirms that the low-complexity algorithm can be used in the extreme scenario. }}

Fig. \ref{fig:pm} shows the effect of the probability of the static state on the accuracy. It can be seen that the proposed scheme offers the same accuracy as the conventional scheme when the probability is low. As we can imagine, the threshold should be set close to 0 when the probability of the static state is low; otherwise, a large number of action instances would be detected as the static state, which reduces the recognition accuracy. In this case, the impact of the action detection module on the ISCC system is small. As the probability of the static state increases, the threshold increases, and the action detection module shows its impact, i.e., reducing the required computation resource. Thus, more computation resource can be allocated to computation tasks, and then more communication resource can be saved for sensing. It eventually gives rise to an increase in sampling rate and accuracy improvement. In the same way, the accuracy of the average communication resource scheme and average computation resource scheme also increases with the probability of the static state. The above results demonstrate that the proposed sensing framework is applicable for the scenario with a high probability of the static state. This conclusion is consistent with Theorem \ref{thm:gain}, which proves that the performance gain brought by the proposed sensing framework is positively related to the probability of the static state. \revision{Moreover, the performance of the proposed low-complexity algorithm is worse than that of the conventional scheme when the probability of the static state is low. It is because the low-complexity is suboptimal and cannot be applied in this case.}

Fig. \ref{fig:nuser} shows the effect of the number of devices on the accuracy. We can observe that the accuracy of the four analyzed schemes all decreases with the number of devices. It is because we need to meet the delay requirements of computation tasks, and thus the communication and computation resources are preferentially allocated to the devices. The remaining amount of resource decreases with the number of devices, which decreases the accuracy. \revision{When the number of devices is 10, the computation resource at the edge server is abundant, and thus the sampling rates of the four schemes are high enough, and the sensing accuracy approaches the upper limit, i.e., 100\%.  When the number of devices is 20, the computation resource at the edge server is limited and thus the sampling rates of the four schemes are low enough, and the sensing accuracy approaches the lower limit, i.e., 12.5\%. } Besides, the accuracy of the four schemes is almost the same when the number of devices is 10 and 20. In contrast, the accuracy of the proposed and conventional schemes drops more slowly than that of the average communication/computation resource scheme. It demonstrates that the proposed resource allocation strategy can suppress the effect of the number of devices.

To show the effectiveness of the threshold selection policy, we also compare the proposed scheme with the fixed threshold scheme under different probabilities, as shown in Fig. \ref{fig:s_eta}. To be specific, we consider that the ratio of threshold to its upper limit  $\eta_u$ is constant in the fixed threshold scheme. From the figure, our proposed scheme shows better performance since the proposed threshold selection policy is optimal, in which the threshold is adaptively selected based on the allocated resource and the probability. Moreover, the threshold that achieves the peak of the fixed threshold scheme increases with the probability of the static state. This is consistent with the explanation for Fig. \ref{fig:pm}, where the threshold increases with the probability of the static state for reducing the unnecessary computation resource cost.

Fig. \ref{fig:tmax} shows the effect of the delay requirement of the sensing task on the accuracy of the four analyzed schemes. From the figure, the impact of the delay requirement is tiny when the delay requirement becomes loose. This result is reasonable since the accuracy grows slowly when the sampling rate is higher than $50$ Hz, as shown in Fig. \ref{fig:acc}. In this case, the decrease of the delay requirement may lead to a drop in the sampling rate but would not lead to a decline in the accuracy. As the delay requirement becomes strict, the sampling rate becomes lower than $50$ Hz. In this case, the accuracy drops rapidly as the delay requirement decreases according to Fig. \ref{fig:acc}. Moreover, the performance of the proposed scheme drops more slowly than the conventional scheme, further demonstrating the proposed scheme's superiority.

\section{Conclusion}

In this paper, we studied an ISCC system and developed an effective sensing framework with an action detection module to avoid unnecessary computation cost for the static state. The sampling theorem was adopted for analyzing the performance of the proposed sensing framework. Through analysis, we identified that the performance gain brought by our proposal is positively related to the probability of the static state. Furthermore, a sensing accuracy maximization problem was formulated with the delay constraints of the sensing task and communication tasks. After analyzing the structure, the original problem was decomposed into two subproblems. By solving them, we proposed an optimal resource allocation strategy and an optimal threshold selection policy and then developed an overall algorithm to maximize the accuracy. Finally, the analyses for the proposed sensing framework and the proposed algorithm were validated by a real-world test, which demonstrated the superiority of our proposals.

\appendices

\section{Proof of Theorem \ref{thm:gain}} \label{proof:thm_gain}
First of all, we should note that the proposed sensing framework is the same as the conventional one when $\eta=0$, i.e., $A=\alpha(F^{\text{s}})$. To obtain the performance gain, we need to find that $p^{\text{CNN}}$ should be less $1$ with $A\ge \alpha(F^{\text{s}})$. Since $p^{\text{CNN}}$ decreases with $\eta$, it is equivalent to find an $\eta >0$ that satisfies  $A\ge \alpha(F^{\text{s}})$. A sufficient condition is that $A$ increases with $\eta$ when $\eta$ approaches $0$, which means $\dfrac{\partial A}{\partial \eta}|_{\eta=0} >0$. Thus, we can obtain the condition in (\ref{eq:gain_condition}) with simple mathematical calculations.

Next, we can calculate the performance gain in two different cases.

1) \emph{Case 1:} If $A$ is always higher than $\alpha(F^{\text{s}})$ when $\eta\in \left(0, \eta_u \right]$, $\eta$ should be $ \eta_u$ for obtaining the highest performance gain. Therefore, $\rho$ is given by
\begin{equation}
	\rho =  1-p^{\text{CNN}}|_{\eta = \eta_u}.
\end{equation} 

2) \emph{Case 2:} If there is an $\eta$ that satisfies  $A=\alpha(F^{\text{s}})$ when $\eta\in \left(0, \eta_u   \right]$, we have the following equation:
\begin{align}
	\alpha(F^{\text{s}}) = A = & \sum_{i=2}^{I }p_i^m\left(1-p_i^o\right)\alpha(F^{\text{s}}) \nonumber \\ 
	&+ p_1^m \left((1-p^l) + p^l\alpha(F^{\text{s}})\right).
\end{align}
Then, $\rho$ is given by
\begin{align}
	\rho &= 1-p^{\text{CNN}} = 1 - \left(\sum_{i=2}^{I}p_i^m\left(1-p_i^o\right)  + p_1^m  p^l\right) \nonumber \\
	&= 1 - \frac{A-p_1^m(1-p^l)}{\alpha(F^{\text{s}})} = \frac{p_1^m (1-p^l)}{A}.
	%1-\frac{T^{\text{s}}}{T^{\text{CNN}}} = \left( \sum_{i=1}^{I-1}p_i^m\left(1-p_i^o\right)  + p_0^m  (1-p^l)  \right) = \frac{p_0^m (1-p^l)}{A}.
\end{align}

Combining both cases, the performance gain is given as shown in (\ref{eq:gain}), which ends the proof.

\section{Proof of Theorem \ref{thm:resource}} \label{proof:thm_convex}
According to the KKT conditions, the necessary and sufficient conditions of the optimal solution can be expressed as 
\begin{align}
	\frac{\partial \mathcal{L}}{\partial \tau_n^{\text{c},\star}} & = - \lambda_n^{\star} \frac{V_n}{\left(\tau_n^{\text{c},\star}\right)^2R_n} + \mu^{\star} = \left\{
	\begin{array}{ll}
		=0, &\tau_n^{\text{c},\star} > 0,\\
		\ge 0, &\tau_n^{\text{c},\star} = 0,
	\end{array}	
	\right. \label{kkt-1}\\
	\frac{\partial \mathcal{L}}{\partial f_n^\star} & = 1 - \lambda_n^{\star} \frac{V_nC_n}{\left(f_n^\star\right)^2}= \left\{
	\begin{array}{ll}
		=0, &f_n^\star > 0,\\
		\ge 0, &f_n^\star = 0,
	\end{array}	
	\right. \label{kkt-2}
\end{align}
%\vspace{-5ex}
\begin{gather}
	\lambda_n^{\star} \left(T_n - T^{\max}_n\right) =0,~~\lambda^{\star} \ge 0,~~n=1,2,\cdots,N,  \label{kkt-3}\\ \mu^{\star}\left(\tau^{\text{s}}F^{\text{s}} + \sum_{n=1}^N \tau^{c,\star}_n - 1\right)=0,~~ \mu^{\star} \ge 0. \label{kkt-4}
\end{gather}
From constraint (\ref{pb_o:t_user}), we can find that $\tau_n^{\text{c},\star} > 0$ and $f_n^\star > 0$. Combining (\ref{kkt-1}) and (\ref{kkt-2}), we can derive the relationship between $f_n^\star$ and $\tau_n^{\text{c},\star} $, as 
\begin{equation}
	f_n^\star = \tau_n^{\text{c},\star} \sqrt{\mu^{\star} R_n C_n},~n=1,2,\cdots,N,
\end{equation}
and we can find that $\mu^\star>0$ and $\lambda_n^{\star}>0$. Next, according to (\ref{kkt-3}), we have
\begin{equation}
	\tau_n^{\text{c},\star} =\dfrac{V_n}{T^{\max}_{n}}\left(\dfrac{1}{R_n}+\sqrt{\dfrac{C_n}{\mu^\star R_n}} \right),~n=1,2,\cdots,N,
\end{equation}
where $\mu^\star$ satisfies $\tau^{\text{s}}F^{\text{s}} + \sum_{n=1}^N \tau^{\text{c},\star}_n =1$. Thus, we can derive that 
\begin{equation}
	\mu^\star=\frac{\sum_{n=1}^N \frac{V_n}{T_n^{\max}} \sqrt{\frac{C_n}{R_n}} }{\left(1-\tau^{\text{s}}F^{\text{s}}-\sum_{n=1}^N\frac{V_n}{T_n^{\max}R_n} \right) },
\end{equation}
which ends the proof.

\section{Proof of Lemma \ref{lem:upper}} \label{proof:lem_up}
We aim to find an upper bound of $F^{\text{s}}$, which is equivalent to find the lower bound of the required communication resource for computation tasks. Therefore, we consider a special case where all devices have the highest data rate, i.e., $\min\limits_n R_n$, and the lowest computation load, i.e., $\left(\min\limits_n V_n, \min\limits_n C_n, \max\limits_n T^{\max}_n\right)$, and all computation resource is allocated to the devices. Then, according to (\ref{eq:T_device}), the devices need less communication resource in this case compared with the optimal solution. The  corresponding required communication resource of computation tasks can be given by
\begin{equation}
  \frac{ N \min\limits_n V_n f^{\text{e}}}{ \min\limits_n R_n \left( \max\limits_n T^{\max}_n f^{\text{e}} -N \min\limits_n V_n \min\limits_n  C_n\right)}.
\end{equation} 
Thus, we can express an upper bound of $F^{\text{s}}$ as (\ref{eq:fs_up}), which ends the proof.

\end{document}